\documentclass[11pt]{article}

\usepackage[margin = 2cm]{geometry}
\usepackage{graphicx}
\usepackage{amssymb}
\usepackage{lineno}
\usepackage[title]{appendix}

\usepackage{multirow}
\usepackage{setspace}
\usepackage{natbib}
\usepackage{booktabs}
\usepackage{caption}
\usepackage{amsmath}
\usepackage{hyperref}
\usepackage{ccicons}
\usepackage[T1]{fontenc}
\usepackage{enumitem}
\usepackage[table,xcdraw]{xcolor}
\usepackage{floatrow}
\usepackage{subcaption}
\usepackage{authblk}
\usepackage{booktabs}
\usepackage[normalem]{ulem}
\useunder{\uline}{\ul}{}
\graphicspath{ {././} }

\begin{document}

\title{A multimodal approach to SME credit scoring integrating transaction and ownership networks \footnote{\scriptsize NOTICE: This is the author’s version of a work which will be submitted for publication. Changes resulting from the publishing process, such as editing, corrections, structural formatting, and other quality control mechanisms may not be reflected in this document. This work is made available under a \href{https://creativecommons.org/licenses/by/4.0/}{Creative Commons BY-NC-ND license}. \ccbyncnd}}

\author[1]{Sahab Zandi}
\author[2,3]{Kamesh Korangi}
\author[4]{Juan C. Moreno-Paredes \footnote{\scriptsize At the time this research was conducted, Dr Moreno-Paredes was with Santander UK plc.}}
\author[5,6]{Mar\'{i}a \'{O}skarsd\'{o}ttir}
\author[2,3]{Christophe Mues}
\author[1]{Cristi\'{a}n Bravo}

\affil[1]{Department of Statistical and Actuarial Sciences, Western University, 1151 Richmond Street, London, Ontario, N6A 5B7, Canada.}
\affil[2]{University of Southampton Business School, University of Southampton, SO17 1BJ, United Kingdom.}
\affil[3]{Centre for Operational Research, Management Sciences and Information Systems, University of Southampton, SO17 1BJ, United Kingdom.}
\affil[4]{Independent Researcher, 18 Hulse Road, Southampton, SO15 2QU, United Kingdom.}
\affil[5]{School of Mathematical Sciences, University of Southampton, SO17 1BJ, United Kingdom.}
\affil[6]{Department of Computer Science, Reykjav\'{i}k University, Menntavegur 1, 102 Reykjav\'{i}k, Iceland.}

\date{}
\maketitle

\begin{abstract}
Small and Medium-sized Enterprises (SMEs) are known to play a vital role in economic growth, employment, and innovation. However, they tend to face significant challenges in accessing credit due to limited financial histories, collateral constraints, and exposure to macroeconomic shocks. These challenges make an accurate credit risk assessment by lenders crucial, particularly since SMEs frequently operate within interconnected firm networks through which default risk can propagate. This paper presents and tests a novel approach for modelling the risk of SME credit, using a unique large data set of SME loans provided by a prominent financial institution. Specifically, our approach employs Graph Neural Networks to predict SME default using multilayer network data derived from common ownership and financial transactions between firms. We show that combining this information with traditional structured data not only improves application scoring performance, but also explicitly models contagion risk between companies. Further analysis shows how the directionality and intensity of these connections influence financial risk contagion, offering a deeper understanding of the underlying processes. Our findings highlight the predictive power of network data, as well as the role of supply chain networks in exposing SMEs to correlated default risk.
\end{abstract}


\section{Introduction}
\label{Introduction}

Small and Medium-sized Enterprises (SMEs) are crucial contributors to global economic growth, innovation, and employment, yet they often face significant barriers in securing the credit necessary to accelerate growth. A primary challenge stems from their status as private entities, resulting in financial disclosures that tend to be more opaque compared to those of larger corporations. This complicates the credit risk assessment process for financial institutions and, therefore, affects both the availability and cost of essential funding for SMEs \citep{bakhtiari2020financial}. Traditionally, the assessment of corporate credit risk relied on financial ratios and historical performance data, with models such as Altman's Z score \citep{altman1968financial} widely used to predict the risk of bankruptcy. However, these models exhibit significant limitations when applied to SMEs, as the latter may lack standardised, audited, or publicly available financial information. This underscores the need for more robust approaches that can better capture the unique risk factors associated with SME lending.

SMEs are embedded in complex supply chain networks, where interconnectivity shapes the dynamics of credit risk \citep{jackson2021systemic}. These networks introduce systemic interactions that affect default risk and financial stability \citep{allen2000financial}, which traditional credit risk models often do not capture \citep{thomas2017credit, long2022clues}. A key factor in this context is correlated default, whereby the probability of one firm defaulting is related to that of another, due to shared economic conditions or sector-specific shocks \citep{nagpal2001measuring}. Ignoring such dependencies can lead to a misjudgment of systemic risk \citep{fenech2015loan}, which shows the necessity to incorporate network-based insights into credit risk models \citep{oskarsdottir2019value, bravo2020evolution}.

In this context, network science has emerged as a powerful tool for analysing such systems, due to its ability to reveal underlying structures and dependencies that traditional methods might overlook \citep{barabasi2016network}. Specifically, SMEs, and the different types of financial and operational interactions between them, can be conveniently represented as multilayer networks \citep{kivela2014multilayer}. The latter are capable of drawing connections from multiple sources without forcing them into a single-layer structure, thus providing a richer and more nuanced view of these interactions. Hence, multilayer networks have proven to be effective in improving our understanding of risk propagation and contagion effects \citep{oskarsdottir2021multilayer}. The diversity of inter-firm relationships raises important questions about which types of connections are most predictive of risk. Moreover, the different layers in a multilayer network may contain either directed or undirected edges, depending on the nature of the underlying relationships. Whilst common ownership relationships are non-directional, there is extensive literature showing that adverse credit risk events or shocks can have ripple effects that extend both upstream and downstream through the supply chain \citep{spatareanu2023bank}, but that the magnitude of those effects may vary in either direction, depending on a variety of other factors \citep{agca2022credit}. Such effects may, among other things, be triggered by supplier-side losses on trade credit obligations or the customer's supply of goods being affected, respectively. Thus, directionality and intensity can play a critical role in contagion dynamics, and ignoring these edge-level properties may obscure asymmetric dependencies. Hence, the ability to fully capture these different types of connection adds to the appeal of multilayer networks in this setting.

For network science to offer valuable insights into these interconnected risks, it needs to be supplemented with advanced modelling techniques that are capable of capturing not just the complexities of SME interactions but also the effects of more conventional, often numeric firm-level risk drivers. The ability to integrate unstructured (or semi-structured) data with structured data represents a major advancement in machine learning \citep{zhang2020combining}. Empirical studies in other fields have shown that combining network data, which is considered semi-structured data, with structured data can indeed help improve the predictive model performance \citep{rao2023fusion}. The concept of multimodal fusion \citep{zhao2024deep}, which involves the merging of diverse information channels (in a way that goes beyond simple concatenation), has been shown to provide more precise predictions than models based solely on unimodal data, in a variety of application settings ranging from sentiment and emotion analysis \citep{poria2017review} to corporate credit risk prediction \citep{tavakoli2023multi, lu2025efficient}. However, most applications in the latter setting have thus far focused on combining textual and numeric data modalities. In contrast, work on multimodal fusion that incorporates network data in the SME context remains scarce and largely unexplored.

Hence, this paper sets out a novel approach that integrates multilayer networks into SME credit risk modelling, explicitly addressing the correlations and contagion risks inherent among interconnected SMEs. Our study focuses on improving the precision of credit risk assessments for SMEs and gaining valuable insights by leveraging the information contained in these networks. Our research questions are thus:

\begin{enumerate}
  \item Does incorporating network data into predictive models improve their accuracy or provide a more in-depth understanding of how credit risk spreads among interconnected SMEs? If so, what is the most effective way to combine network data with structured data?

  \item Is it advantageous to use multilayer networks over single-layer networks, and which type of connection between SMEs proves to be more informative?

  \item What insights can be gained by considering the directionality and intensity of these connections?
\end{enumerate}

The remainder of the paper is structured as follows. The next section reviews prior work on machine learning with network data and SME credit risk modelling. Section~\ref{Methodology} details the methodology, including multilayer networks and the models created for this study. Section~\ref{Experimental setup} discusses the data, network construction, and experiments conducted. Section~\ref{Results} presents the experimental results. Section~\ref{Discussion} highlights some discussion points and summarises our findings. Finally, the last section provides conclusions and suggests directions for future research.

\section{Previous work}
\label{Previous work}

\subsection{Modelling credit risk for SMEs using network data}
\label{Modelling credit risk for SMEs using network data}
Acknowledging that SMEs are susceptible to the financial distress of related businesses \citep{long2022clues}, a growing body of literature has sought to use network data to enhance credit risk modelling, by incorporating information on connections with suppliers, customers, financial institutions, or ownership ties. This approach is based on network theory, with \citet{borgatti2011network} demonstrating how the position of an organisation within a network can provide critical insights into its financial stability and performance. The objective has been to capture the mutual dependencies among firms and to better understand how financial distress propagates throughout an interconnected ecosystem \citep{berloco2021predicting}.

In a financial network setting, \citet{iori2008network} examined overnight interbank lending, showing that centrality metrics indicate the systemic importance of a node. Extending this logic to SMEs, other studies suggested that the risk of a firm is affected by the financial health of its connected counterparts. \citet{giesecke2004cyclical} provided evidence that cyclical correlations and contagion between firms can amplify portfolio losses, whereas \citet{beaver2019group} showed that group affiliation --- such as parent–subsidiary relationships --- affects default risk through risk sharing within corporate structures. More recently, common ownership has emerged as another channel of contagion. \citet{massa2017information} showed that the presence of ownership links between firms influences credit conditions by altering monitoring incentives. \citet{zhou2022identifying} further found that firms are more likely to exhibit discreditable behaviour if they are thus tied to other discreditable firms, highlighting how reputational risks propagate along ownership structures. Another line of work derives inter-firm links from payment or transaction data. For example, \citet{letizia2019corporate} employed company payment networks and found evidence of risk homophily, as companies with similar credit ratings exhibited a higher probability of being connected. This finding shows how the composition of the neighbourhood relates to creditworthiness. More generally, these studies illustrate how financial links, ownership ties, and transactional networks each represent different channels of contagion --- though most have focused on one type of connection at a time.

Several empirical studies have also sought to operationalise relational risk by extracting statistical features from inter-firm networks. \citet{vinciotti2019effect} analysed how the number of financially linked firms and their default history influence the risk of SME credit, distilling relational information into a series of neighbour counts and binary indicators. Although informative, such aggregated measures simplify the heterogeneous financial health of counterparties into coarse indicators, limiting their ability to capture the dynamics of distress propagation. Other work has taken this further by modelling correlated default dependencies \citep{calabrese2019birds} or employing multilayer network structures \citep{oskarsdottir2021multilayer} to quantify contagion channels. Yet, such approaches tend to be assumption-driven, relying on pre-specified models of how dependencies or layers operate, rather than learning flexible relational representations directly from data.

In response to these limitations, this study develops a multimodal network learning framework that integrates both the structural position of SMEs within multilayer networks and the financial attributes of their connected firms. This design moves beyond pre-specified dependencies and aggregated neighbour statistics, instead embedding relational signals in a data-driven manner. The resulting framework provides a more fine-grained and dynamic representation of how inter-firm dependencies influence credit risk, enabling richer insights into the mechanisms of financial contagion in SME networks.

\subsection{Modelling credit risk for SMEs using multimodal learning}
\label{Modelling credit risk for SMEs using multimodal learning}
The primary challenge in assessing SME credit risk lies in the diversity and informality of SME operations. Financial data on SMEs is often limited, making traditional credit scoring methods less effective. Furthermore, SMEs exhibit a high degree of heterogeneity in their operations, which financial ratios alone cannot fully capture \citep{smit2012literature}. Advances in artificial intelligence have sparked interest in applying multimodal learning techniques \citep{ramachandram2017deep} to enhance the accuracy and reliability of credit risk assessments for small companies. The rationale for adopting multimodal learning is to integrate complementary information from diverse sources while reducing redundancy due to overlapping features across modalities. By allowing each modality to be represented with its own parameters, the framework preserves modality-specific information and avoids the oversimplification that can arise from forcing different data sources through a single shared representation \citep{baltruvsaitis2018multimodal}.

In SME credit risk assessment, multimodal learning looks beyond financial ratios to include unstructured data sources such as social media engagement, online reviews, and news coverage. For example, \citet{stevenson2021value} employed a multimodal approach to predict small business default, incorporating both standard financial data and textual evaluations from loan officers. \citet{mai2019deep} showed how the combination of textual data with traditional accounting-based ratios and market-based variables improved the prediction of corporate bankruptcy. \citet{zhang2022credit} fused firm-level demographic attributes -- such as age, size, industry, and geographic location -- with behavioural financial data -- such as payment histories, credit use, and transaction records -- to assess the risk of SME credit in supply chain finance. Using accounting, market, and pricing data, \citet{korangi2023transformer} introduced a multimodal transformer-based framework to predict default probabilities for mid-cap companies. \citet{tavakoli2023multi} developed multimodal models to predict external credit ratings by applying different fusion strategies to textual data originating from earnings call transcripts and numerical data. 

Incorporating a network data modality into SME credit risk models represents another step forward, enabling a more comprehensive view of creditworthiness by considering the interdependencies that affect firm performance. Specifically, this study advances SME credit risk modelling by leveraging explicit inter-firm network information. Using multimodal learning, we derive novel insights into SME default risk from two distinct data modalities. Unlike previous studies that merged these heterogeneous data into a single modality \citep{vinciotti2019effect, yin2020evaluating, rishehchi2021data}, we explicitly treat them as separate inputs, allowing independent learning of their respective contributions and combined representation through data fusion. This yields a richer depiction of SME credit risk, capturing how firm characteristics and inter-firm connections jointly influence the probability of default. Importantly, embedding the network structure directly into the risk representation also provides a lens to infer the mechanisms of financial contagion.

\section{Methodology}
\label{Methodology}

In this section, we describe our methodology. First, we detail the process of building multilayer networks. Next, we describe the different types of graph neural networks (GNNs) that we use to encode topological dependencies in the networks. Following this, we elaborate on information fusion strategies and then describe the different models implemented in this study and their respective architectures.

\subsection{Multilayer networks}
\label{Multilayer networks}
\begin{figure}[tb]
\includegraphics[scale=0.9]{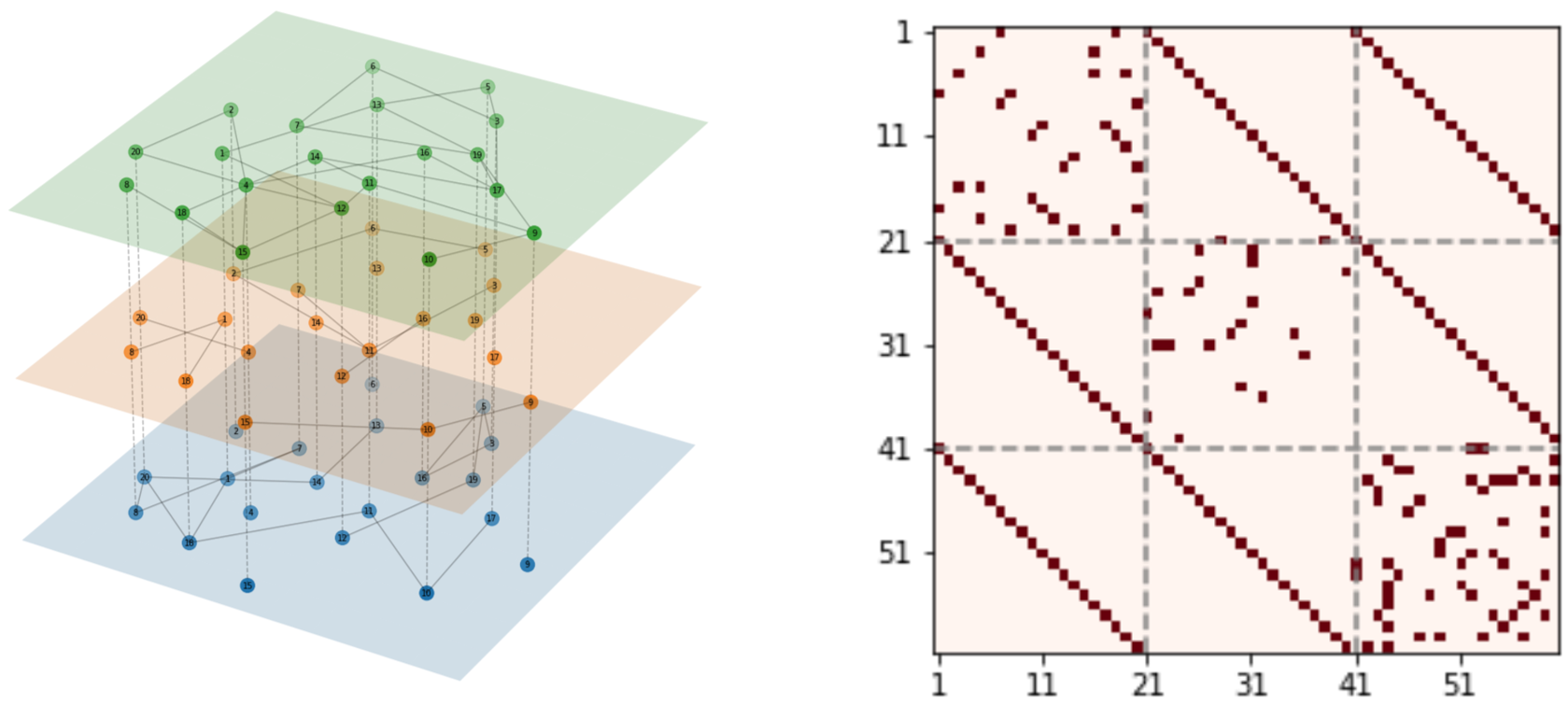}
\centering
\caption{A multilayer network (left) and its supra adjacency matrix (right). Available under the CC-BY license from \citet{zandi2024attention}.}
\label{fig1}
\end{figure}

Consider a network represented as $G=(V,A,X)$, where $V=\{v_1, v_2,...,v_n\}$ denotes the collection of nodes, $n=|V|$ represents the total number of unique nodes and $X\in\mathbb{R}^{n \times d}$ is a feature matrix. In this matrix, $X_i$ is a column vector that contains the characteristics of node $v_i$, and $d$ is the number of features. The network is characterised by its supra-adjacency matrix, $A\in\mathbb{R}^{nl \times nl}$, where $l$ denotes the number of layers. This matrix records the connectivity between pairs of nodes within the same layer and between pairs of nodes across different layers, that is, if $v_i$ from layer $k$ and $v_j$ from layer $m$ are connected $(1\leq k,m\leq l)$, then $A_{(k-1)n+i,(m-1)n+j}=1$; otherwise, the value is $0$. In such a network, although every layer shares the same node set, the connections between them differ. Each layer is dedicated to a specific type of relationship, with edges within a layer linking nodes based on relatedness. In addition, a series of interlayer edges simply specify which nodes are identical. To dive deeper into the network's structure, an embedding derived from the network is necessary, a process detailed in the subsequent subsection. Fig.~\ref{fig1} illustrates a multilayer network along with its supra-adjacency matrix.

\subsection{Graph neural networks}
\label{Graph neural networks}
GNNs extend traditional neural networks to effectively process and learn from network-structured data. A distinct challenge in working with such networks is capturing the topological dependencies, as neighbouring nodes can influence each other. In this work, we trial two types of GNNs: GAT \citep{velivckovic2018graph} and GIN \citep{xu2018powerful}. These models are employed to extract topological relationships between a node and its neighbours and encode both the network's structural information and the node features, thereby effectively capturing the information embedded in node connections.

\subsubsection{Graph attention networks}
\label{Graph attention networks}
GAT is applied to each $G$ to obtain a hidden representation matrix $Z$. Each row of $Z$ contains a node embedding, which means that for node $v_i$ we have an embedding $Z_i$. The core component of GAT is an attention mechanism that computes the importance of each neighbour to a node. The attention coefficients are computed as follows:
\begin{eqnarray}
&\alpha_{ij}=\text{softmax}_j\left(\text{LeakyReLU}\left(a^{T}[Wh_i||Wh_j]\right)\right).\label{eq1}
\end{eqnarray}

Here, $h_i$ and $h_j$ are the embeddings of nodes $v_i$ and $v_j$, respectively, obtained from the previous layer (or initialised as the raw feature vectors $X_i$ and $X_j$ for the first layer). $W \in \mathbb{R}^{D \times d}$ is a learnable weight matrix where $D$ is the embedding dimension, $a^T \in \mathbb{R}^{1 \times 2D}$ is a learnable weight vector, \(\Vert\) denotes concatenation, and $\alpha_{ij}$ represents the attention coefficient indicating the importance of the features of node $v_j$ relative to node $v_i$. For each node, the GAT aggregates features from itself and its neighbours weighted by the attention coefficients. The updated embedding for node $v_i$ is calculated as:
\begin{eqnarray}
&Z_i=\sum_{{v_j}\in N(v_i)\cup\{v_i\}}\alpha_{ij}Wh_j.\label{eq2}
\end{eqnarray}

Here, $Z_i$ represents the new embedding of node $v_i$ after one GAT layer and $N(v_i)$ denotes the neighbours of node $v_i$. The embeddings are iteratively updated in each layer. Note that we also include the self-edge for each node. GAT often employs multihead attention for more stable learning. This means running several independent attention mechanisms in parallel and concatenating (or averaging) their outputs. 

GATs are powerful because they dynamically assign importance to nodes' neighbours, enabling the model to focus on the most relevant information in the network. This adaptability makes GATs effective for various network-based tasks where the structure and connectivity of the data are crucial.

\subsubsection{Graph isomorphism networks}
\label{Graph isomorphism networks}
GIN, a second type of GNN used in this study, was developed to address the limitations of the previous GNN in distinguishing between different network structures. The key concept behind GIN is to enable the model to have the same discriminative power as the Weisfeiler-Lehman (WL) network isomorphism test \citep{weisfeiler1968reduction}, a classical algorithm used for network comparison. In GIN, each node aggregates information from its neighbours to update its own features. The update mechanism is designed to be as powerful as the WL test in terms of distinguishing different network structures. The hidden representation of each node is updated based on both its own embedding and the aggregated embeddings of its neighbours. The update rule is formulated as follows:
\begin{eqnarray}
&Z_i=\text{MLP}\left((1+\varepsilon)h_i+\sum_{j\in N(v_i)}h_j\right).\label{eq3}
\end{eqnarray}

Here, \(\varepsilon\) is a learnable parameter or a fixed scalar, and MLP represents a multilayer perceptron. As before, the embeddings $h_i$ and $h_j$ are iteratively updated in each layer, with $h_i = X_i$ as the initial node representation in the first layer. The update rule ensures that a node's own embedding is considered in addition to the aggregated embeddings of its neighbours. The inclusion of \(\varepsilon\) allows the model to learn the importance of a node's own embedding relative to its neighbours. 

The strength of GIN lies in its ability to effectively capture the structural information of networks. It can distinguish between different network structures that other GNN models might fail to differentiate. This makes GIN particularly useful in tasks that require a detailed understanding of network topology, such as social network analysis, where embedding the precise structure of the network is crucial.

\subsection{Information fusion}
\label{Information fusion}
To enable deep learning from both structured and semi-structured data, we will examine the effectiveness of different information fusion approaches. 

\subsubsection{Fusion levels}
\label{Fusion levels}
\begin{figure}[tb]
\includegraphics[scale=0.6]{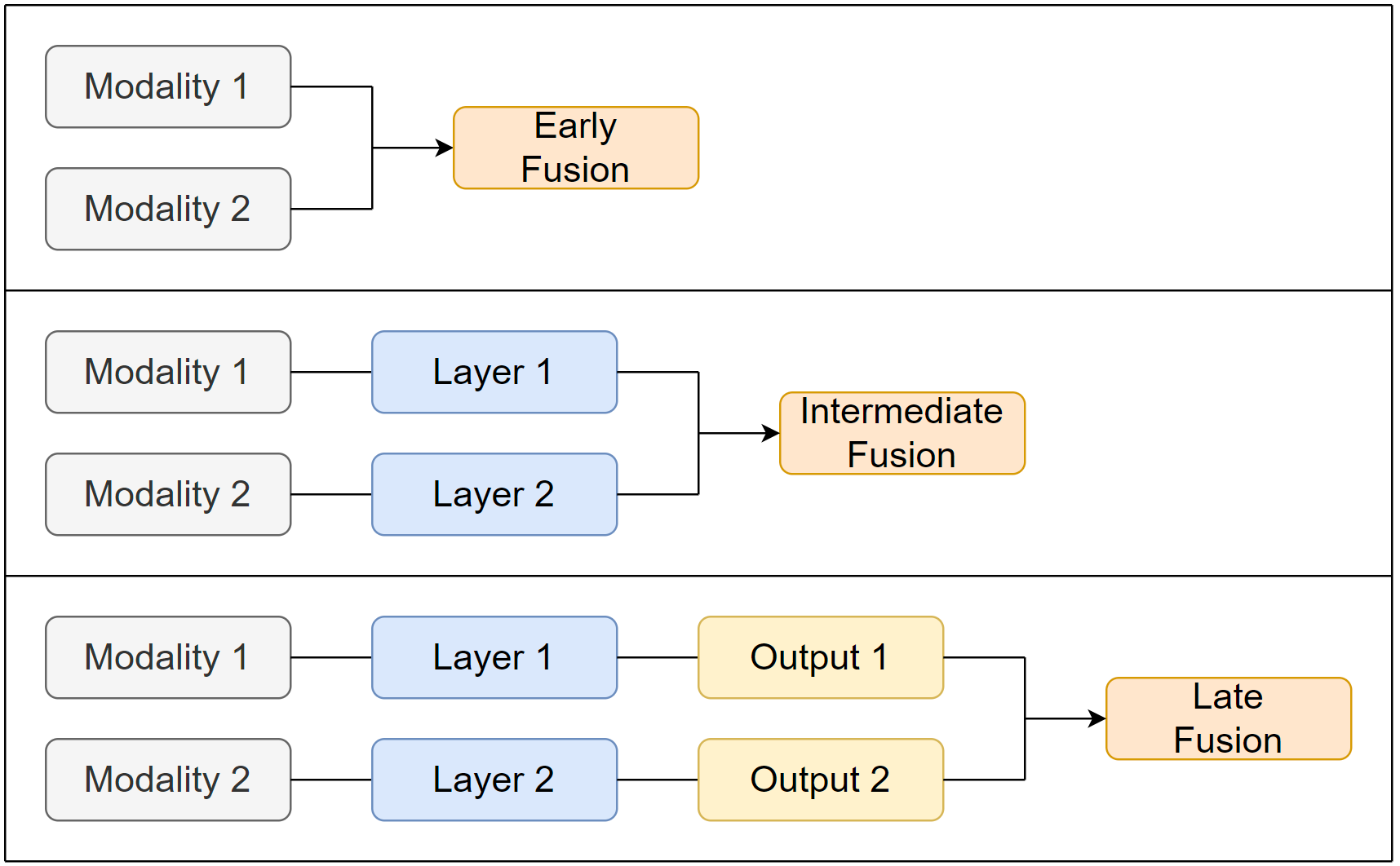}
\centering
\caption{The three primary fusion levels: early, intermediate, and late.}
\label{fig2}
\end{figure}

Information fusion refers to the association, correlation, and combination of data from single or multiple sources \citep{zhang2018multi}. The concept of fusion level refers to the specific point in the processing pipeline where this merging occurs. Fig.~\ref{fig2} shows the three primary fusion levels: early, intermediate, and late fusion. Early fusion, also known as signal fusion, merges raw data directly with the aim of retaining maximal information and leaving filtering to the model. By avoiding manual feature selection or aggregation, it minimises bias by not imposing prior assumptions about feature relevance. However, the resulting high-dimensional feature space can increase variance if not properly regularised \citep{luo1988multisensor}. Intermediate fusion is a higher-level fusion in which data merging does not occur until after initial processing by earlier layers of the deep learning model \citep{boulahia2021early}. Late fusion, the highest level, involves the integration of information that has already been extracted as abstract features or that can even assume the form of predictions generated by previous models \citep{luo1988multisensor}. Additionally, hybrid fusion combines different fusion levels, such as early and late fusion, when integrating information from different modalities \citep{chango2022review}. In this study, we consider both simple and hybrid fusion, employing early and intermediate fusion levels. By combining both levels, we seek to leverage their respective benefits, as early fusion can capture a broad range of unprocessed information, whilst intermediate fusion handles more refined features.

\subsubsection{Fusion techniques}
\label{Fusion techniques}
This study explores two fusion techniques: concatenation and cross-attention. The former is a simple operation-based method, while the latter is an attention-driven approach \citep{zhang2020multimodal}. Concatenation as a data fusion strategy involves combining feature vectors from multiple modalities into one extended feature vector. This new vector then serves as an input for machine learning tasks. The simplicity of concatenation makes it widely applicable, although it does not account for complex interactions between features from different modalities.

The second method, that is, cross attention, is a more intricate fusion technique that is especially relevant in deep learning settings where understanding the interplay between different types of data is crucial \citep{lee2018stacked}. For two different modalities, the model generates queries $(Q)$ from one modality and keys $(K)$ and values $(V)$ from the other modality. Attention scores are calculated by taking the dot product of queries with keys. This is often scaled down by the square root of the dimensionality of the keys to avoid overly large dot product values, which can lead to gradient vanishing problems during training. The raw scores are then normalised across all keys using the \textit{softmax} function, which turns them into a distribution of weights that sum to one. These attention weights are subsequently used to create a weighted sum of the values, which gives us the final output of the cross-attention mechanism, emphasising the parts of one modality that are most relevant to each element of the other modality. The formula for this calculation is as follows:

\begin{eqnarray}
&\text{Output}=\text{softmax}(\frac{QK^T}{\sqrt{d_k}})V.\label{eq4}
\end{eqnarray}

Here, $d_k$ is the dimensionality of the keys and $K^T$ represents the transpose of the key matrix. The output of the cross-attention mechanism can then be used as a fused representation of the two modalities, combining the information in a way that is contextually enriched by the intermodal relationships. Cross-attention thus allows the model to adaptively focus on the most pertinent information from one modality informed by another.

\subsubsection{Fusion strategies}
\label{Fusion strategies}
\begin{figure}[tb]
\centering
\begin{subfigure}[b]{0.5\textwidth}
\includegraphics[scale=0.35]{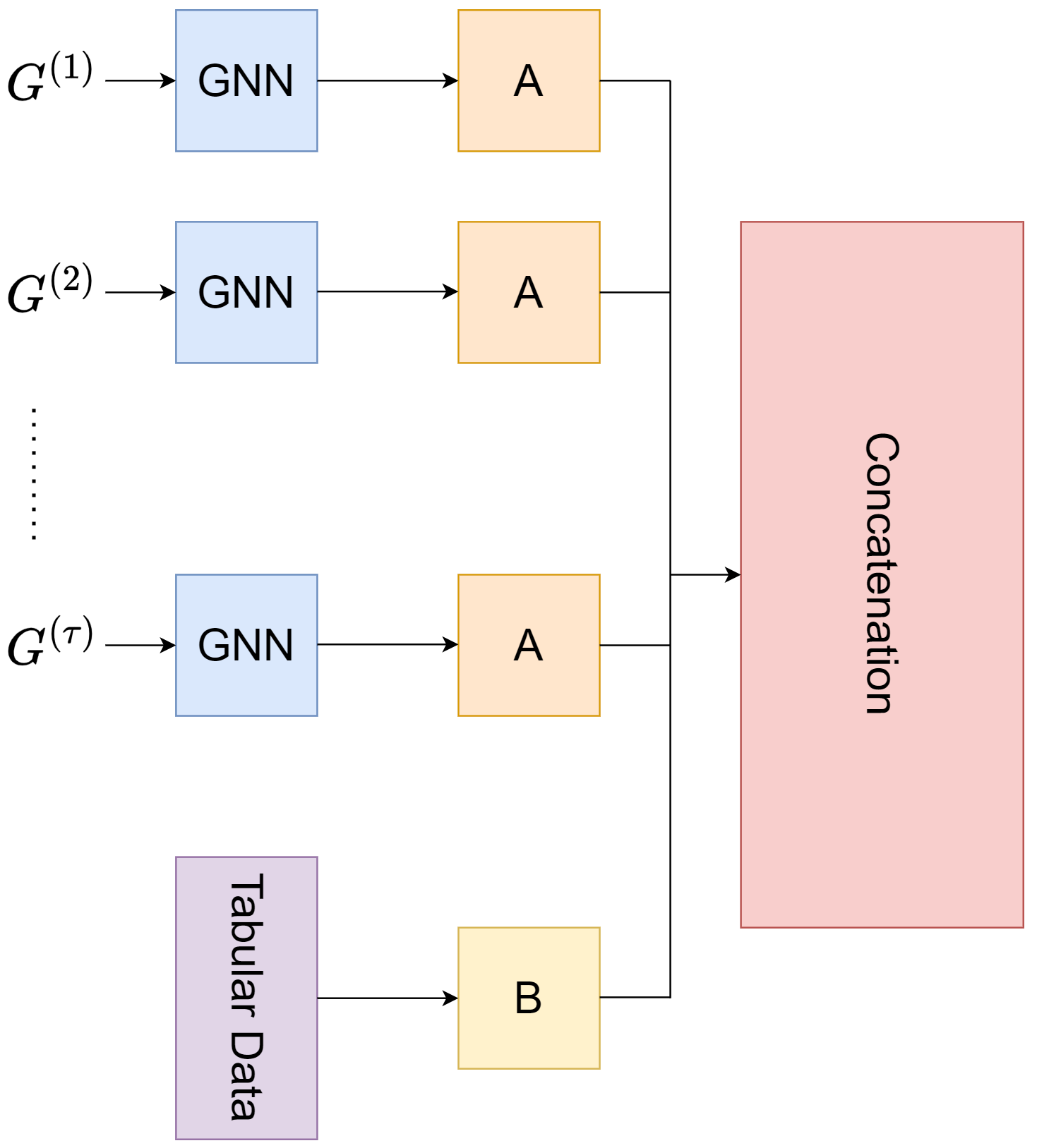}
\caption{Strategy 1 - Simple Concatenation}
\label{fig3a}
\end{subfigure}
\quad
\begin{subfigure}[b]{0.45\textwidth}
\includegraphics[scale=0.35]{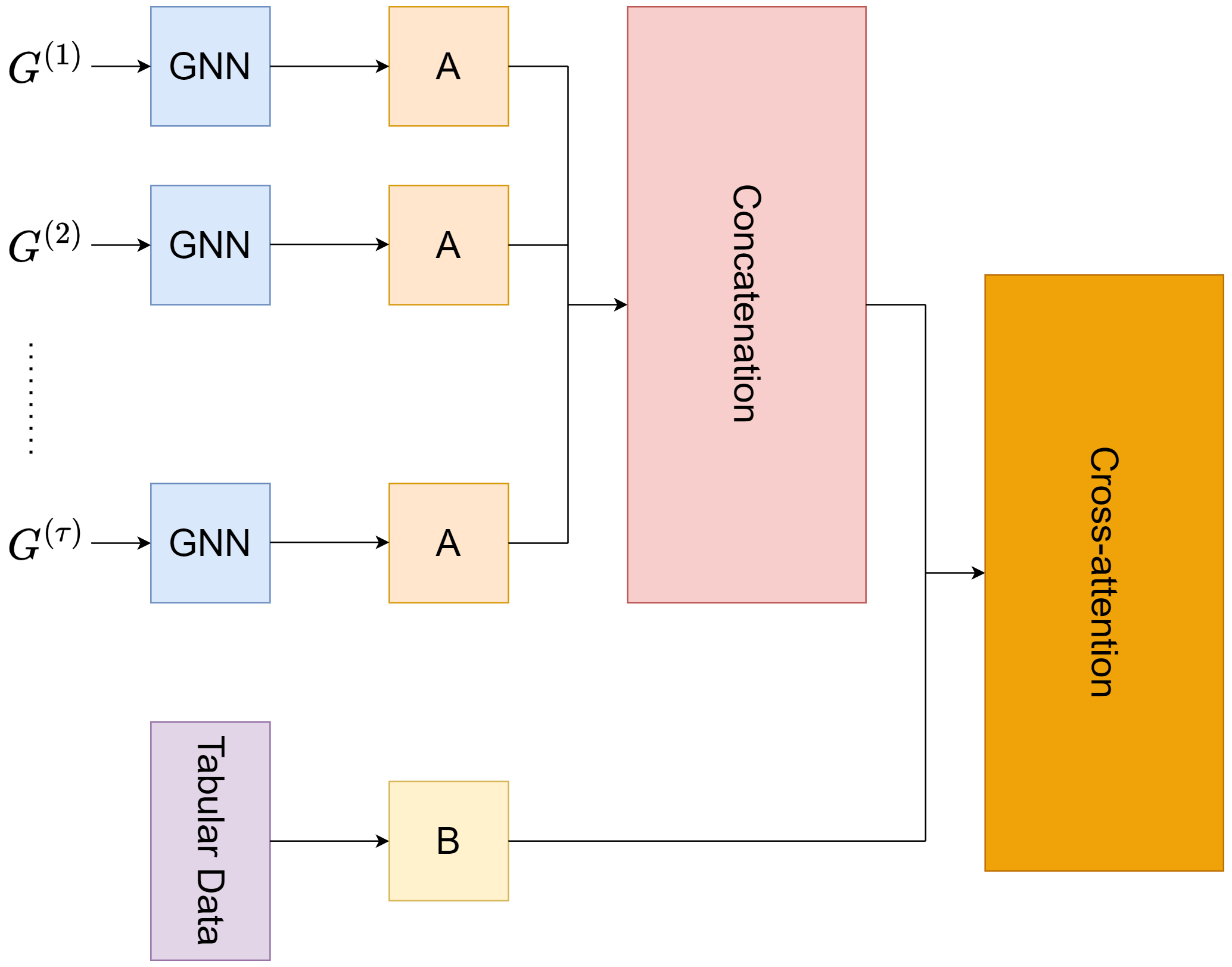}
\caption{Strategy 2 - Simple Concatenation-Attention}
\label{fig3b}
\end{subfigure}
\newline
\newline
\newline
\begin{subfigure}[b]{0.5\textwidth}
\includegraphics[scale=0.35]{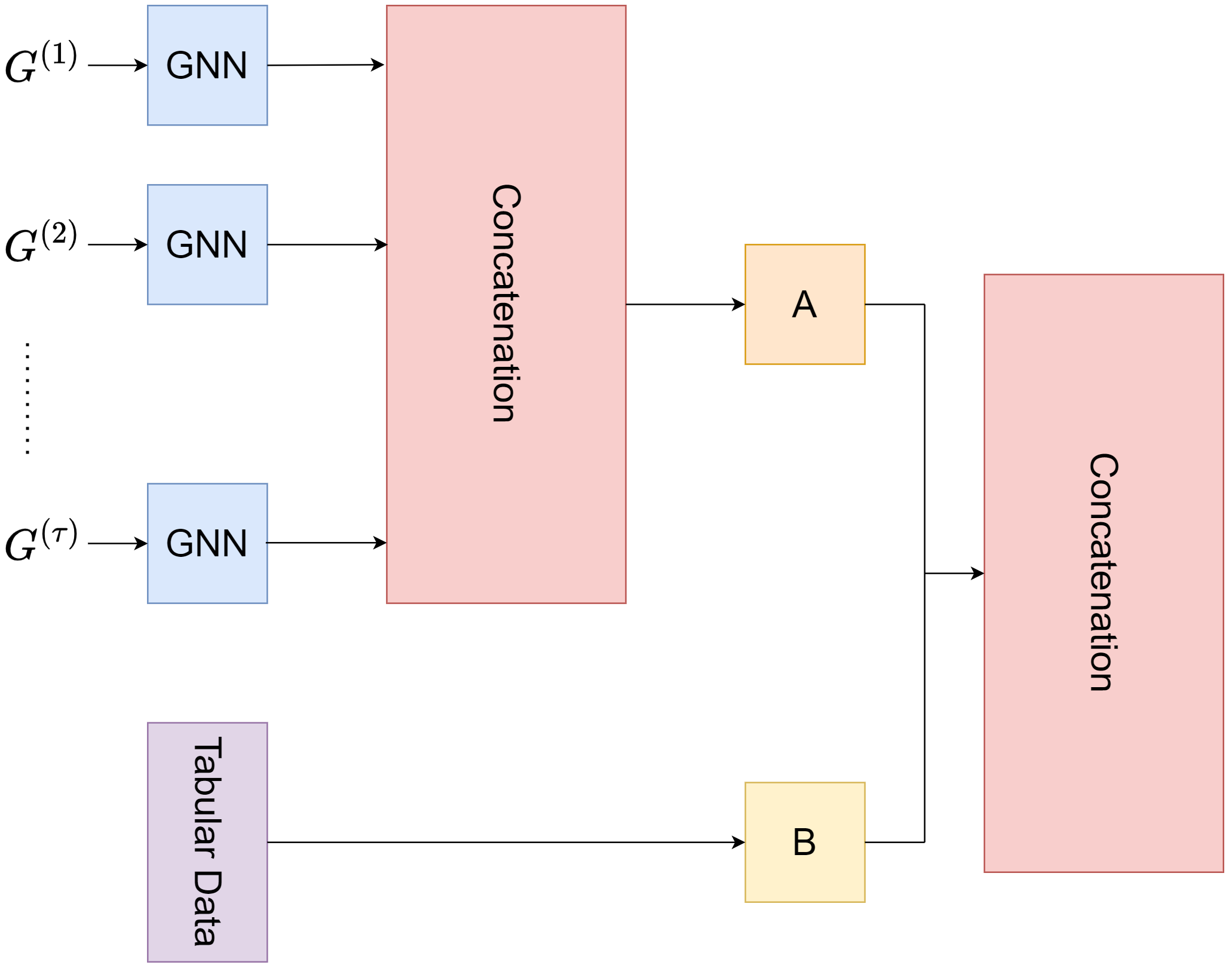}
\caption{Strategy 3 - Hybrid Concatenation}
\label{fig3c}
\end{subfigure}
\quad
\begin{subfigure}[b]{0.45\textwidth}
\includegraphics[scale=0.35]{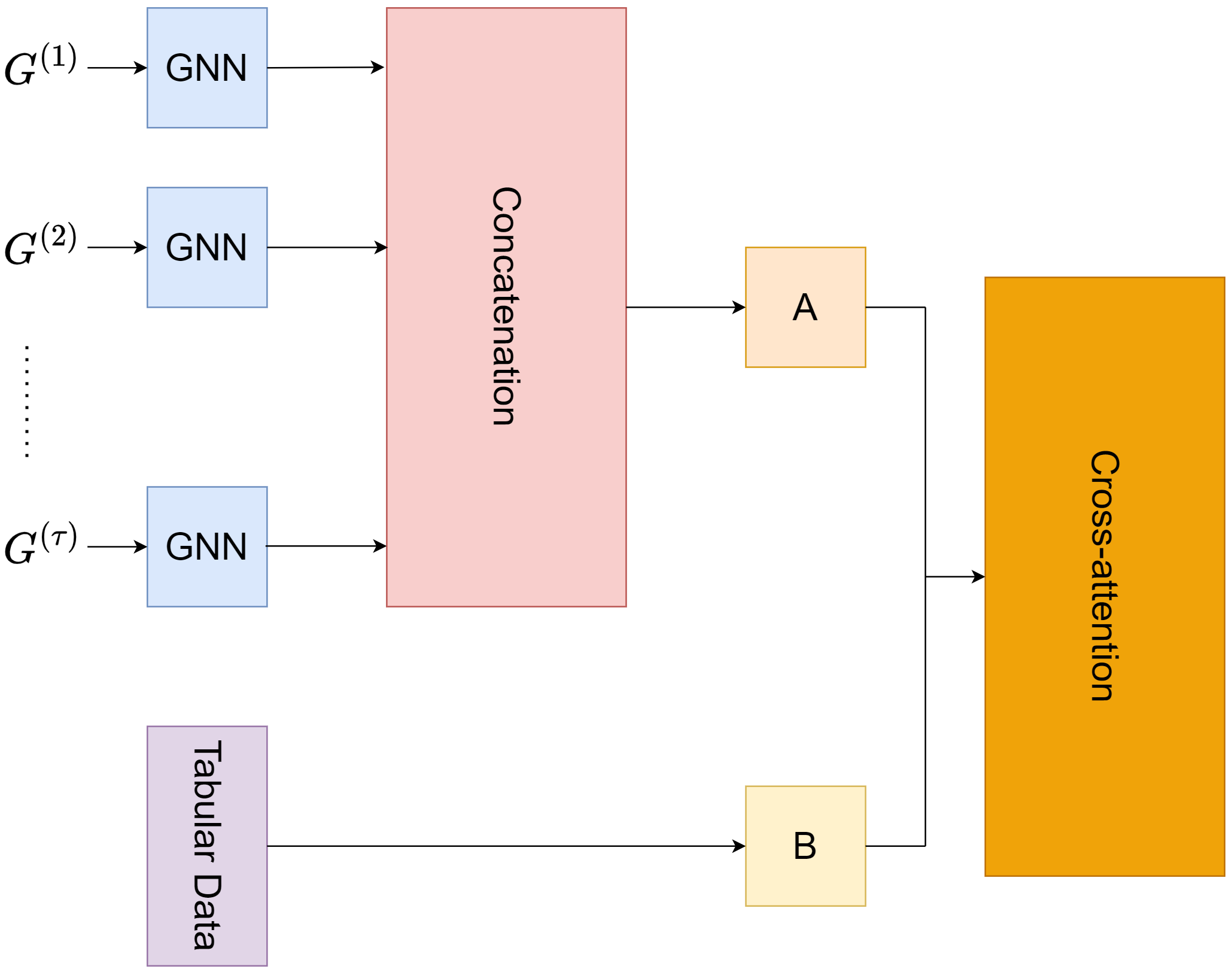}
\caption{Strategy 4 - Hybrid Concatenation-Attention}
\label{fig3d}
\end{subfigure}
\caption{The proposed fusion strategies for the multimodal models. Adapted from \citet{tavakoli2023multi}.}
\label{fig3}
\end{figure}

Our empirical analysis hence aims to identify, for our specific application, the most effective combination of: 1) GNN type, 2) fusion 
techniques, and 3) fusion levels. In our experiments, we test both GAT and GIN (applied to a sequence of $\tau$ graph snapshots, as Section \ref{Network construction} will later explain), each of which we combine with four different fusion strategies \citep[as illustrated in Fig.~\ref{fig3} adapted from][]{tavakoli2023multi}. This creates a total of eight multimodal model variations, designed to deal with two types of input channels: the semi-structured (network) data and the structured (tabular) data (hence, the resulting models will be referred to as bimodal models). The respective fusion strategies are classified as follows.

\begin{itemize}
  \item Strategy 1 -- Simple Concatenation: As shown in Fig.~\ref{fig3a}, the semi-structured and structured channels are processed through the (neural) networks A and B, respectively. Their penultimate layers are concatenated employing a simple form of intermediate fusion.

  \item Strategy 2 -- Simple Concatenation-Attention: Fig.~\ref{fig3b} shows a setup in which the penultimate layers of the semi-structured modality are concatenated and then fused using cross-attention with the neural network layer for the structured data.

  \item Strategy 3 -- Hybrid Concatenation (see Fig.~\ref{fig3c}): This setup uses a hybrid fusion approach, in which all semi-structured data undergo early fusion and are fed into neural network A. The structured data is processed through network B. The penultimate layers of the different modalities are then merged by concatenation.

  \item Strategy 4 -- Hybrid Concatenation-Attention (Fig.~\ref{fig3d}): This strategy is similar to the previous strategy, except that it uses cross-attention to fuse the penultimate layers of the two modalities.  
\end{itemize}

\begin{figure}[hbt!]
\includegraphics[scale=0.7]{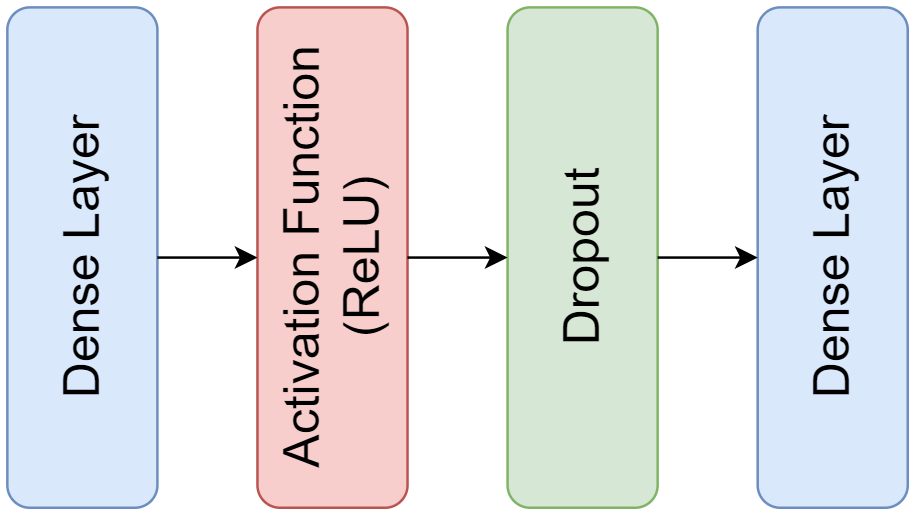}
\centering
\caption{An overview of the architectures of networks A and B.}
\label{fig4}
\end{figure}

In Fig.~\ref{fig3}, networks A and B consist of a set of densely connected neurons followed by a series of layers that either apply a chosen activation function to introduce non-linearity or use a dropout function for regularisation. The parameters for each of these networks, such as neuron count and dropout rate, are optimised for the best performance on the validation set. Fig.~\ref{fig4} provides an overview of the architectures of these networks.

\subsection{Unimodal and bimodal models}
\label{Unimodal and bimodal models}
\begin{figure}[tb]
\centering
\begin{subfigure}[b]{0.60\textwidth}
\includegraphics[scale=0.45]{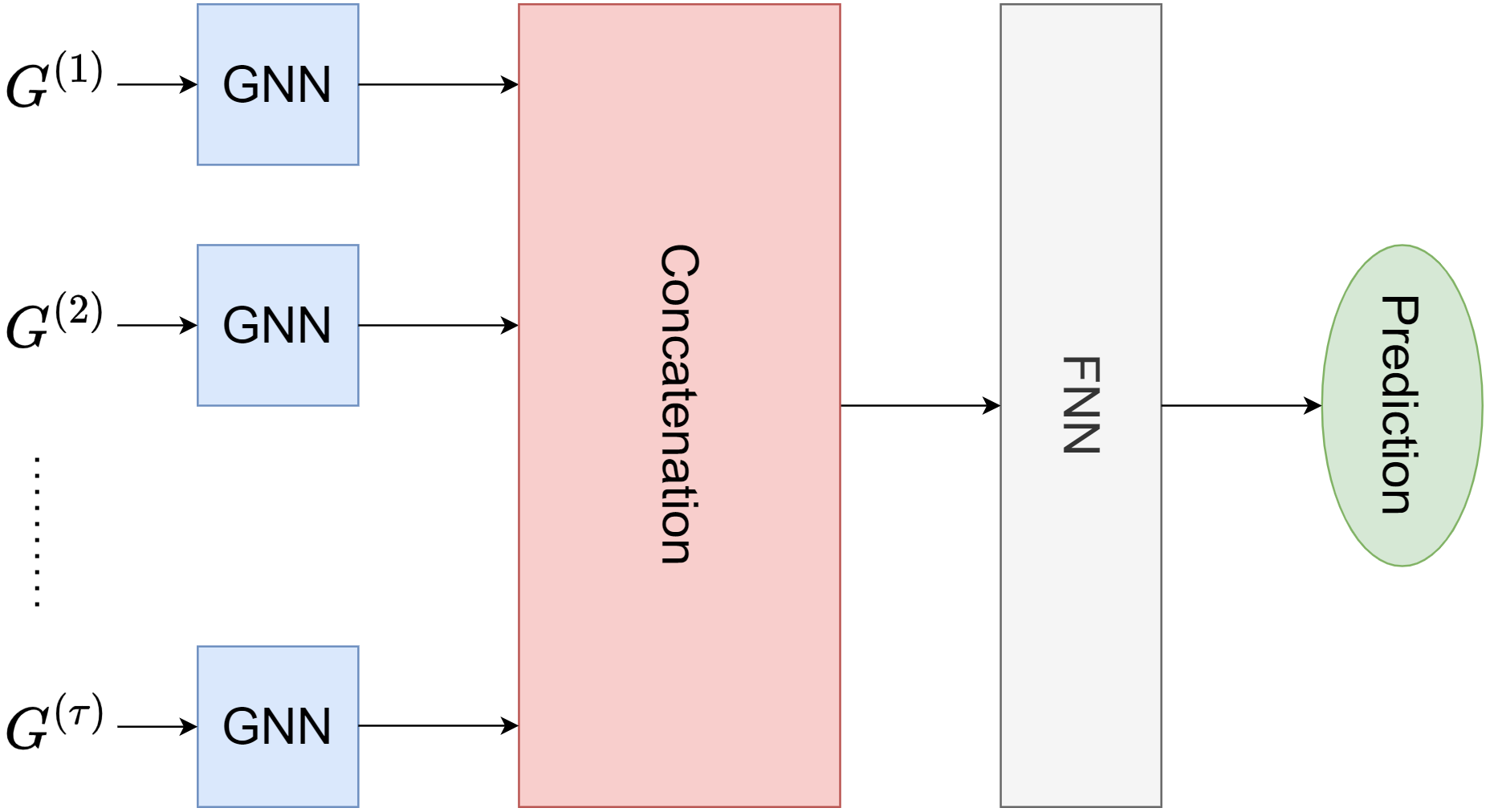}
\caption{Unimod-GNN}
\label{fig5a}
\end{subfigure}
\begin{subfigure}[b]{0.38\textwidth}
\includegraphics[scale=0.34]{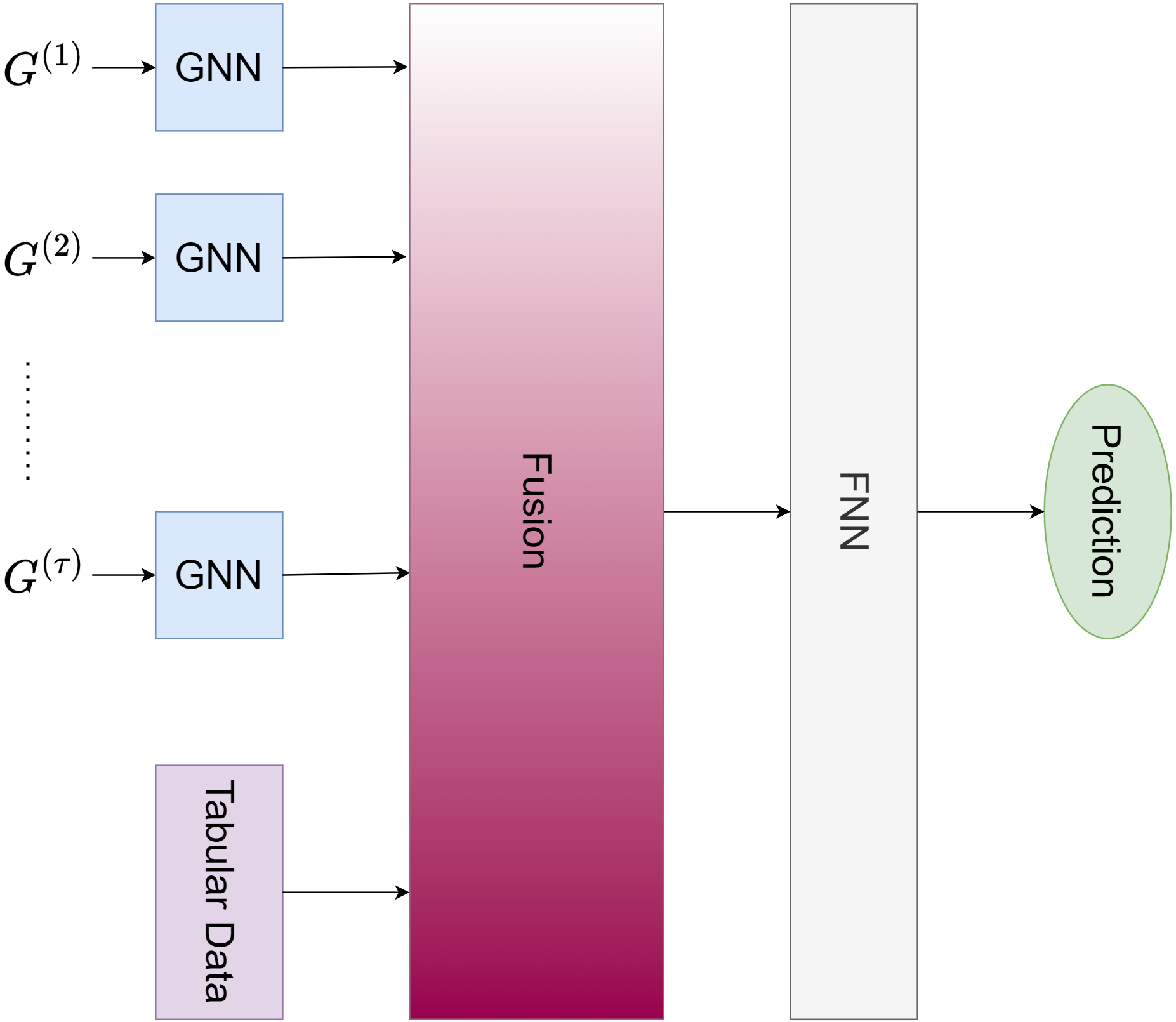}
\caption{Bimod-GNN}
\label{fig5b}
\end{subfigure}
\caption{The unimodal and bimodal model structures.}
\label{fig5}
\end{figure}

In addition to our proposed bimodal model variations, we also tested some simpler unimodal alternatives. Referred to as Unimod-GNN, these are single-channel (unimodal) models, built using either GAT or GIN, that exclusively employ the network data and within which each network node contains a self-loop. By connecting each node back to itself, we aim to preserve the node's own features during the learning process. Fig.~\ref{fig5a} presents a schematic representation of this model, comprising a sequence of processes that begin with multiple GNN instances being used to process the different network snapshots. Next, the outputs of these GNNs are concatenated into a single representation. The combined data is then fed into a feedforward neural network (FNN), which further processes the information to produce a final prediction output.

The second series of models, i.e.\@ Bimod-GNN, are two-channel (bimodal) models that process the network and the tabular data, alongside each other. Fig.~\ref{fig5b} shows their structure. As before, a series of GNNs are used to process the network data, but now their output is fused with the tabular data using one of the strategies described in the previous subsection. The outputs from the fusion process then again go through an FNN which allows for further processing before arriving at the final output.

\subsection{FNN and loss function}
\label{FNN and loss function}
Although the architecture of the FNN in Fig.~\ref{fig5} exhibits similarities in terms of layout and design with the structure of networks A and B in Fig.~\ref{fig4}, a notable difference in the FNN architecture is the incorporation of an additional layer toward the end, which employs a sigmoid activation function. The purpose of this final layer is to refine the model output to make binary decisions, in this case determining whether a specific node (loan) $v_i$ should be classified into class $1$ ($Y_i=1$) or class $0$ ($Y_i=0$). We let $\hat{Y}_i$ denote the probability that node $v_i$ belongs to class $i$ produced by the model.

A critical element in training and evaluating the performance of a deep learning model is the choice of loss function. In the context of this study, we employ the well-known binary cross-entropy loss function \citep{gneiting2007strictly}, which can be written as:
\begin{equation}
\textrm{Loss}=-\frac{1}{n}\sum_{i=1}^{n} \left[Y_ilog(\hat{Y}_i)+(1-Y_i)log(1-\hat{Y}_i) \right].\label{eq5}
\end{equation}

\section{Experimental setup}
\label{Experimental setup}

\subsection{Dataset}
\label{Dataset}
In this paper, we use a comprehensive loan-level dataset provided by a prominent financial institution, containing information on loans issued to SMEs. The dataset includes data on both the loan and the borrower, covering approximately 218,000 companies and 1.2 million loan instances. Within the data, a single company may have been granted multiple loans. The features are available at the time of loan application and do not change from one month to the next. Numerical features are normalised with min-max scaling. We clean the data by treating outliers and handling null values. Specifically, outliers are capped at the 99\textsuperscript{th} percentile and 1\textsuperscript{st} percentile points. To handle null values in categorical and numerical features, we implement a strategy comparable to {\citeauthor{bravo2013granting}}'s (\citeyear{bravo2013granting}) (summarised in Appendix~\ref{Strategy for handling null values}). In addition, to reduce redundancy and improve model performance, we remove features with high correlation, using a correlation threshold of 70\%. When deciding between correlated features, we retain the one that is most strongly correlated with the target. After completing the data preparation process, we are thus left with 65 distinct features per loan that will be used to predict one-year-ahead loan default. The descriptions of the most informative features are shown in Table~\ref{tab1}.

\begin{table}[hbt!]
\small
\caption{Description of most informative node features.}
\begin{center}
\begin{tabular}{l|l}
\toprule
Feature & Description\\
\midrule
avg\textunderscore bal\textunderscore lia & Average balance of liabilities over the past 12 months\\
mon\textunderscore install\textunderscore amt & Monthly instalment amount for existing mortgage loans\\
tot\textunderscore rev\textunderscore gen & Total revenue generated from operations of active assets\\
if\textunderscore act\textunderscore flag\textunderscore na & Is the active flag indicator for the entity set to `N/A'?\\
end\textunderscore cur\textunderscore year & Endowment of the entity for the current year\\
avg\textunderscore bal\textunderscore act & Average balance of active assets\\
if\textunderscore pre\textunderscore app & Does the entity have pre-approved status?\\
ent\textunderscore own\textunderscore equ & Entity's own equity\\
cov\textunderscore rat\textunderscore tax & Coverage ratio before taxes\\
if\textunderscore desc\textunderscore gsi & Is the description of the current worst GSI situation available?\\
adj\textunderscore tot\textunderscore rev\textunderscore gen & Adjusted total revenue generated from operations associated with liabilities\\
if\textunderscore wel\textunderscore dep\textunderscore acc & Does the offer involve a welcome deposit account with the bank?\\
if\textunderscore restruct & Has the entity undergone any restructuring?\\
if\textunderscore bur\textunderscore ind & Is there any bureau indicator assigned to the entity?\\
if\textunderscore trans\textunderscore ind & Is there any transaction indicator assigned to the entity?\\
default & Being 90 days or more in payment arrears over next 12 months (\emph{target})\\
\bottomrule
\end{tabular}
\end{center}
\vspace{-12pt}
\label{tab1}
\end{table}

Loans originated from July 2019 to December 2020 are used for training. For this cohort, we observe a default rate of 3.63\%, providing a suitable environment to train our models. The subsequent test phase is conducted with loans originating from January 2021 to October 2021. A key step in our experimental setup is the exclusion of any companies from the test set whose loans were included in the training dataset. This deliberate exclusion is crucial for maintaining the integrity of our test results, ensuring that there is no data overlap that could artificially inflate the performance of our predictive models. 

\begin{figure}[tb]
\includegraphics[scale=0.2]{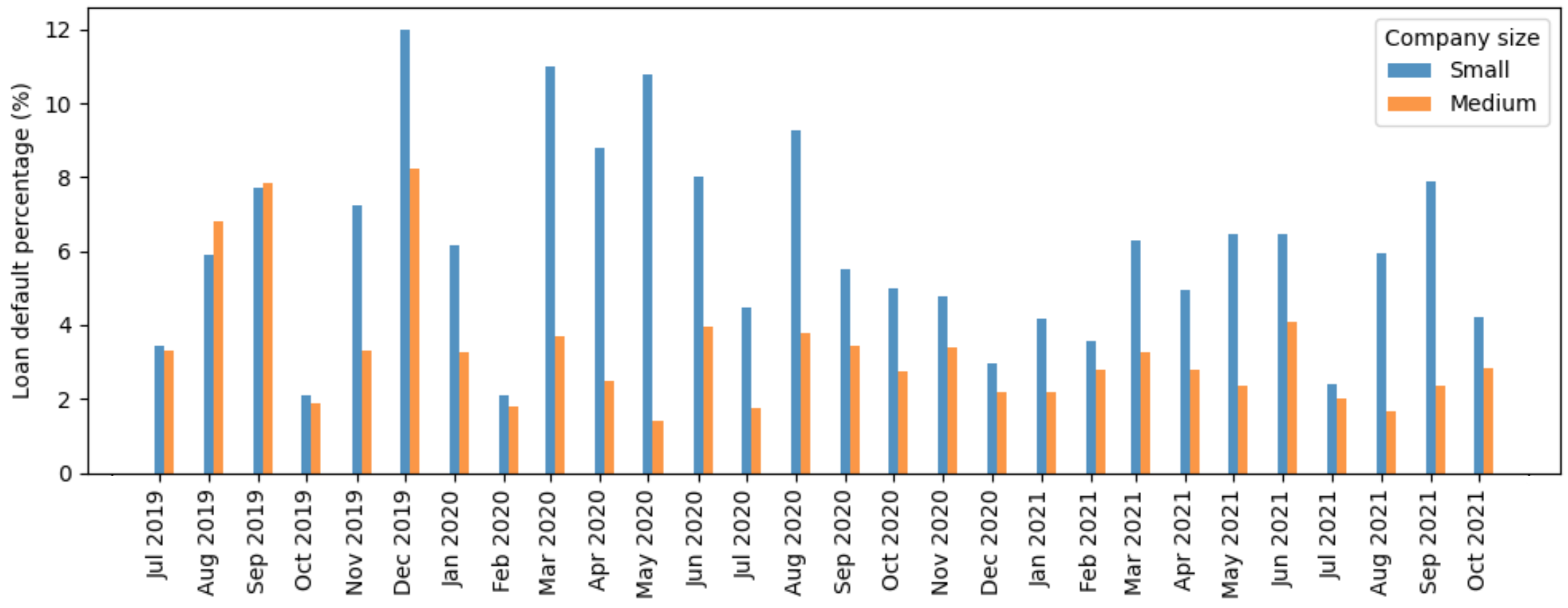}
\centering
\caption{Loan default rates by company size and loan origination month.}
\label{fig6}
\end{figure}

Fig.~\ref{fig6} presents the loan default rates categorised by company size (small or medium) and loan origination month. It reveals a clear trend whereby small companies exhibit higher default rates compared to medium-sized ones. This pattern highlights that smaller companies tend to be more financially vulnerable, possibly due to limited access to resources or higher sensitivity to economic fluctuations.

\subsection{Network construction}
\label{Network construction}

As the original dataset also contains company ownership data and any financial transactions between companies that were made through the lender, we are able to construct single- and double-layer networks (see Subsection~\ref{Multilayer networks}), using financial transactions (FT) and common ownership (CO) as the primary sources of connections. This allows us to map and analyse the complex interplay between different companies. In these networks, the nodes represent loans associated with the companies.

\begin{figure}[tb]
\includegraphics[scale=0.35]{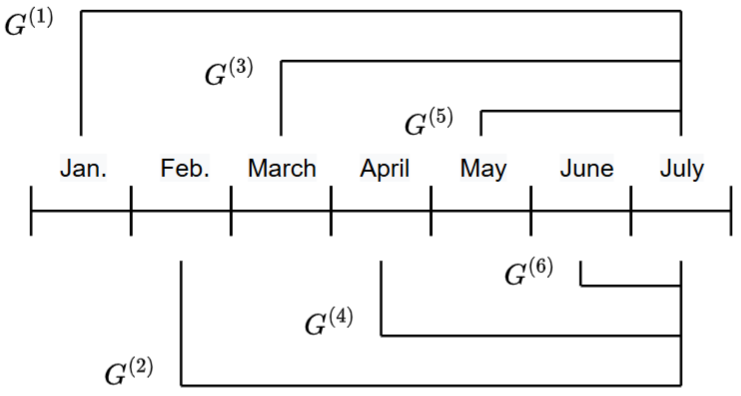}
\centering
\caption{Snapshots corresponding to the July 2019 origination month.}
\label{fig7}
\end{figure}

For each origination month, we adopt a six-month look-back period, which is similar to the approach by \citet{zandi2024attention} but applied in a different context. Specifically, each origination month is assigned six snapshots, each of which represents the connections observed between the loans of that origination month and other loans originating in one of the six preceding months. To illustrate, consider loans that originated in July 2019 (see Fig.~\ref{fig7}). We begin by identifying the companies to which these loans were issued. Next, we observe what other companies exhibited a relationship with those companies at any time from January to June 2019, referring to the former as our pool of `neighbouring' companies. We then establish edges between each of the loans originating in July 2019, and any loans issued to its neighbouring companies, provided that they originated in January 2019. This produces a first snapshot, $G^{(1)}$. To form $G^{(2)}$, we connect loans originating in July 2019 with loans from neighbouring companies originating in February 2019. We continue this process until we connect loans originating in July 2019 with loans from neighbouring companies originating in June 2019, thus forming our sixth snapshot, $G^{(6)}$. This procedure is repeated by moving the origination month by one month each time, ensuring that the entire period is covered.

The rationale for this design is that the effects of inter-firm relationships on credit risk are rarely immediate, but instead materialise over several months through mechanisms such as delayed payments, supply chain disruptions, or gradual transmission of financial distress \citep{berloco2021predicting}. Restricting the look-back period to six months ensures that the connected firms also have relatively recent loan applications in the dataset. This window allows us to capture both short-term and lagged dependencies while avoiding the noise and dilution associated with excessively long windows, thereby striking a balance between economic intuition and predictive tractability in modelling SME credit risk \citep{huang2024time}.

During the training phase for each origination month, our objective is to predict defaults within the 12 months following that month. A one-year horizon is practical for credit management and decision making, offering a balance between providing enough time to accurately assess risk and avoiding the uncertainty associated with longer-term predictions \citep{lopez2000evaluating}.

\subsection{Model evaluation}
\label{Model evaluation}

Our research involves evaluating the predictive performance of several unimodal and bimodal models and comparing them with that of established baseline methods. Given the class imbalance within this binary classification problem, we employ the Area Under the Receiver Operating Characteristic Curve (AUC), which reflects a model’s ability to distinguish between default and non-default cases across the range of thresholds, and the Area Under the Precision–Recall Curve (AUCPR), which emphasises performance on the minority (default) class, as the two primary means of measuring each model's performance. The results of these evaluations are reported, accompanied by 95\% confidence intervals obtained by bootstrapping applied to the test dataset.

We perform a modality contribution analysis to infer the relative importance of each input modality in shaping the fused representation learnt by the model. We also investigate how accounting for the directionality and weight of connections affects the performance of the best-performing model and enhances the analysis of risk propagation. Since, similarly to other operational research problems, explainability is a key consideration \citep{de2023explainable}, we complete the analysis by employing the Shapley approach \citep{lundberg2017unified} to gain additional information on the contributions of individual features.

\section{Results}
\label{Results}

\subsection{Performance of the baseline methods}
\label{Performance of the baseline methods}
To compare our proposed network models, we have trained several baseline models that use only tabular data. Table~\ref{tab2} displays the performance results for four baseline methods: Logistic Regression (LR) \citep{hosmer2013applied}, Random Forest (RF) \citep{breiman2001random}, XGBoost (XGB) \citep{chen2016xgboost}, and a Deep Neural Network (DNN) \citep{lecun2015deep}. Among these, LR is preferred in some settings for its simplicity and ease of interpretation. RF, on the other hand, is known for its robustness and ability to handle large and complex datasets with high accuracy. XGB is recognised as a powerful technique for classification and regression tasks involving structured datasets. In recent years, DNN has gained popularity for a variety of predictive tasks. We perform a grid search to tune the hyperparameters for each model, using a validation set that covers 20\% of the data. Details of tuning and DNN architecture are provided in Appendix~\ref{Hyperparameter tuning for baseline models} and Appendix~\ref{Architecture of the DNN baseline model}, respectively.

\vspace{+6pt}
\begin{table}[hbt!]
\small
\centering
\caption{Performance of the baseline models.}
\begin{tabular}{l|cc}
\toprule
Model & \multicolumn{1}{c}{AUC} & \multicolumn{1}{c}{AUCPR}\\
\midrule
LR & $0.821\pm0.016$ & $0.322\pm0.014$\\
RF & $0.853\pm0.015$ & $0.382\pm0.012$\\
XGB & $\mathbf{0.863\pm0.014}$ & $\mathbf{0.398\pm0.012}$\\
DNN & $0.839\pm0.014$ & $0.345\pm0.010$\\
\bottomrule
\end{tabular}
\label{tab2}
\end{table}

Table~\ref{tab2} reports the out-of-time performance of these baseline models, showing that XGB performs the best among them (indicated in bold), closely followed by RF. Of the four, LR is the weakest performer, which was expected given its linear nature. The DNN's performance sits in between, indicating that while deep learning techniques have potential, their performance can be hindered by the need for extensive tuning and large amounts of data. In general, the baseline results set a high bar against which the proposed models will be compared.

\subsection{Performance of the proposed models}
\label{Performance of the proposed models}
We apply the unimodal and bimodal models proposed earlier to each type of single network layer, i.e.\@ derived either from the financial transaction (FT) or common ownership (CO) data, and to the double-layer networks. To facilitate comparison, the training, validation, and test sets are kept consistent with those used in the baseline model experiments. Details regarding the hyperparameter tuning for GAT and GIN are provided in Appendix~\ref{Hyperparameter tuning for GAT and GIN models}. Next, Table~\ref{tab3} presents the performance of the unimodal models, while Table~\ref{tab4} shows the performance of the bimodal models. The best-performing model in each table is again highlighted in bold.

\vspace{+6pt}
\begin{table}[hbt!]
\scriptsize
\centering
\caption{Performance of the unimodal models.}
\begin{tabular}{lc|cc|cc|cc}
\toprule
\multicolumn{2}{c|}{Model}& \multicolumn{2}{c}{Single Layer: FT} & \multicolumn{2}{|c}{Single Layer: CO} & \multicolumn{2}{|c}{Double Layer: FT-CO}\\
Strategy&GNN & \multicolumn{1}{c}{AUC} & \multicolumn{1}{c}{AUCPR} & \multicolumn{1}{|c}{AUC} & \multicolumn{1}{c}{AUCPR} & \multicolumn{1}{|c}{AUC} & \multicolumn{1}{c}{AUCPR}\\
\midrule
\multirow{2}{*}{Unimod}&GAT & $0.820\pm0.014$ & $0.310\pm0.011$ & $0.813\pm0.012$ & $0.306\pm0.013$ & $\mathbf{0.828\pm0.015}$ & $\mathbf{0.323\pm0.010}$\\
&GIN & $0.814\pm0.014$ & $0.308\pm0.010$ & $0.802\pm0.013$ & $0.282\pm0.012$ & $0.822\pm0.012$ & $0.314\pm0.010$\\
\bottomrule
\end{tabular}
\label{tab3}
\end{table}

For the unimodal models, it can be seen from Table~\ref{tab3} that GAT yields better results than GIN, across all network configurations. The observed performance differential is likely attributable to practical factors such as sensitivity to hyperparameter tuning, training stability, and the risk of overfitting. Moreover, GAT’s attention mechanism provides a useful means of encoding how different neighbours contribute to prediction, which may explain why its representations appear better aligned with the structure of SME networks in our experiments. The best-performing unimodal model, i.e.\@ Unimod-GAT, is still outperformed, though, by the best-performing baseline model, XGB. This outcome is consistent with the known robustness of tree-based ensemble methods for structured tabular data, and underscores that deep learning models are not necessarily superior in any context \citep{borisov2022deep, gunnarsson2021deep}.

\vspace{+6pt}
\begin{table}[hbt!]
\scriptsize
\centering
\caption{Performance of the bimodal models.}
\begin{tabular}{lc|cc|cc|cc}
\toprule
\multicolumn{2}{c|}{Model}& \multicolumn{2}{c|}{Single Layer: FT} & \multicolumn{2}{c|}{Single Layer: CO} & \multicolumn{2}{c}{Double Layer: FT-CO}\\
Strategy&GNN & \multicolumn{1}{c}{AUC} & \multicolumn{1}{c|}{AUCPR} & \multicolumn{1}{c}{AUC} & \multicolumn{1}{c|}{AUCPR} & \multicolumn{1}{c}{AUC} & \multicolumn{1}{c}{AUCPR}\\
\midrule
\multirow{2}{*}{Simple Concat}&GAT & $0.845\pm0.010$ & $0.350\pm0.008$ & $0.841\pm0.012$ & $0.349\pm0.009$ & $0.849\pm0.012$ & $0.361\pm0.007$\\
&GIN & $0.841\pm0.011$ & $0.348\pm0.009$ & $0.840\pm0.011$ & $0.347\pm0.010$ & $0.844\pm0.009$ & $0.358\pm0.008$\\
\multirow{2}{*}{Simple Concat-Att}&GAT & $0.852\pm0.012$ & $0.359\pm0.008$ & $0.844\pm0.011$ & $0.357\pm0.010$ & $0.855\pm0.011$ & $0.366\pm0.009$\\
&GIN & $0.840\pm0.009$ & $0.345\pm0.009$ & $0.839\pm0.012$ & $0.341\pm0.008$ & $0.842\pm0.008$ & $0.355\pm0.008$\\
\multirow{2}{*}{Hybrid Concat}&GAT & $0.857\pm0.012$ & $0.372\pm0.009$ & $0.852\pm0.010$ & $0.365\pm0.008$ & $0.863\pm0.012$ & $0.392\pm0.010$\\
&GIN & $0.853\pm0.010$ & $0.369\pm0.010$ & $0.847\pm0.008$ & $0.357\pm0.008$ & $0.859\pm0.011$ & $0.373\pm0.007$\\
\multirow{2}{*}{Hybrid Concat-Att}&GAT & $0.866\pm0.012$ & $0.397\pm0.011$ & $0.861\pm0.009$ & $0.375\pm0.007$ & $\mathbf{0.872\pm0.008}$ & $\mathbf{0.406\pm0.008}$\\
&GIN & $0.849\pm0.011$ & $0.359\pm0.009$ & $0.845\pm0.009$ & $0.356\pm0.008$ & $0.856\pm0.012$ & $0.371\pm0.010$\\
\bottomrule
\end{tabular}
\label{tab4}
\end{table}

However, the results shown in Table~\ref{tab4} demonstrate that bimodal models generally produce better results compared to unimodal models, while the reported metrics also have narrower confidence intervals, suggesting greater robustness. This improved performance can possibly be attributed to the ability of bimodal models to integrate and leverage information from both the network and traditional structured data simultaneously. We also find that hybrid fusion strategies outperform simple concatenation ones. The latter is likely due to hybrid approaches being capable of preserving detailed signals while also modelling more abstract cross-modal dependencies, which may be useful when combining heterogeneous data such as tabular data and network embeddings. Specifically, the hybrid concatenation-attention strategy using GAT, applied to double-layer networks, achieves the highest AUC of 0.872 and an AUCPR of 0.406, thus outperforming the best baseline model. This underscores the effectiveness of advanced fusion techniques, such as cross-attention, in capturing complex interactions between different modalities. However, it should be noted that there is a slight overlap between the confidence interval of the best-performing model and those of its closest competitors.

Another notable aspect is that the financial transaction network data appears more informative than the common ownership network. Direct financial transactions between firms can point to inter-firm dependencies, providing insight into financial health and stability that prove important for assessing creditworthiness. In contrast, the common ownership network, while valuable, captures more static structural relationships and may not reflect current financial realities as effectively. Ownership links can signal financial support or risk contagion, but lack the detailed transactional information provided by financial transfers.

The results also highlight the predictive performance gains of using double-layer networks over single-layer ones, suggesting that incorporating multiple types of connections allows the models to capture more nuanced patterns of credit risk contagion.

\subsection{Modality contribution analysis}
\label{Modality contribution analysis}
To better understand the relative importance of each modality in the best performing proposed model, that is, the hybrid Concat-Att-GAT, we perform a modality contribution analysis. We compute two separate cross-attention outputs: one where the tabular representation attends to the network representation and another where the network representation attends to the tabular representation.

Formally, let $Q_T$ and $Q_N$ be the queries derived from the tabular and network modalities, respectively. Let $K_T, V_T$ denote the keys and values of the tabular modality and $K_N, V_N$ those of the network modality. We compute two attended representations as follows:
\begin{eqnarray}
&R_N = \textrm{Attn}(Q_T, K_N, V_N), \label{eq6}\\[10pt]
&R_T = \textrm{Attn}(Q_N, K_T, V_T). \label{eq7}
\end{eqnarray}

Here, $R_N$ captures how much the network contributes to the refinement of the tabular query, while $R_T$ captures how much the tabular modality contributes to the representation of the network. We quantify the contribution of each modality using the relative $\ell_2$ norm of the attended outputs as follows:
\begin{eqnarray}
&C_N = \frac{\|R_N\|_2}{\|R_T\|_2 + \|R_N\|_2 + \epsilon}, \label{eq8}\\[10pt] 
&C_T = \frac{\|R_T\|_2}{\|R_T\|_2 + \|R_N\|_2 + \epsilon}. \label{eq9}
\end{eqnarray}

\begin{figure}[tb]
\includegraphics[scale=0.25]{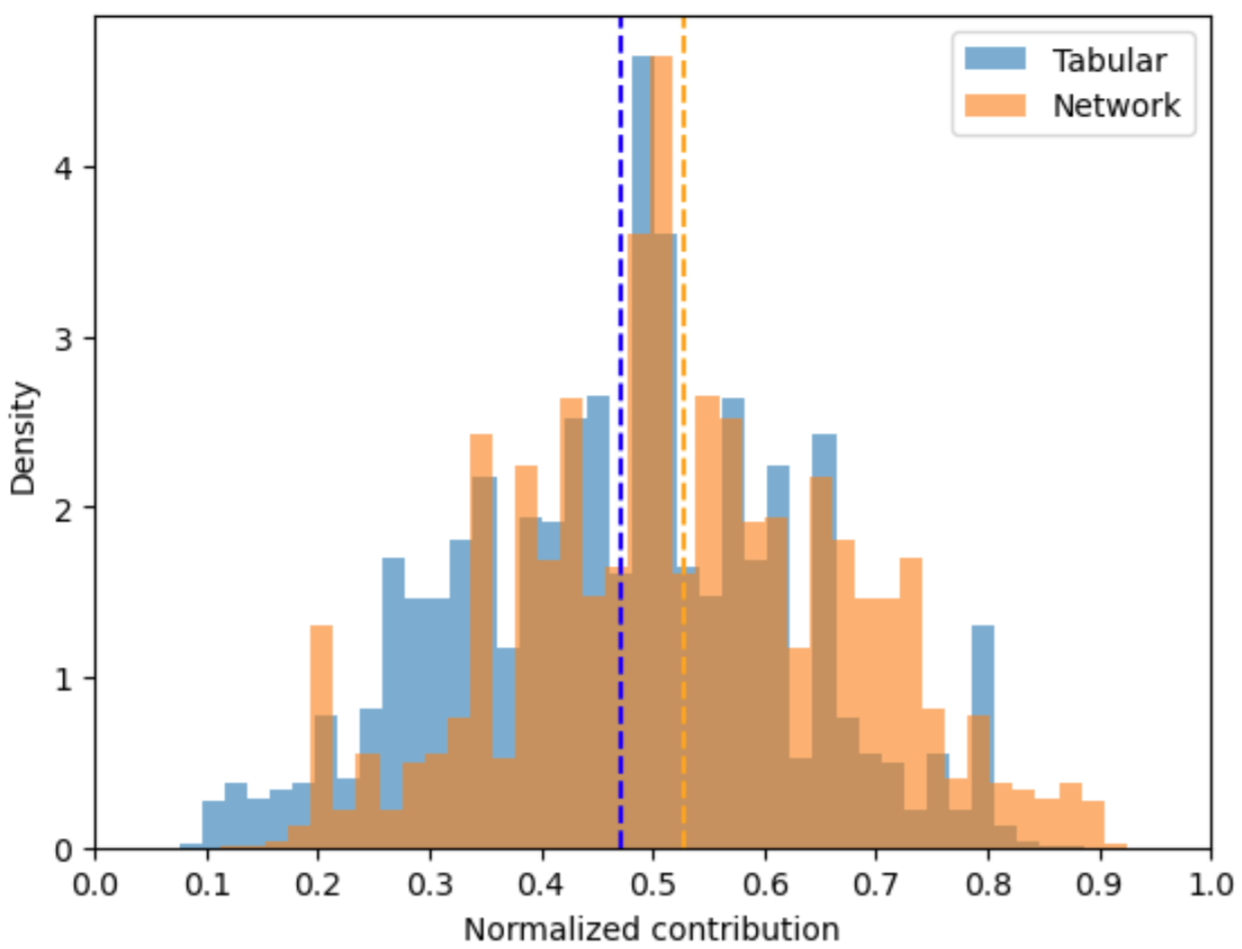}
\centering
\caption{Distributions of normalised modality contributions to the final fused representation. The dashed lines indicate mean contributions from each modality.}
\label{fig8}
\end{figure}

Above, $C_N$ and $C_T$ represent the contributions of the network and tabular modalities, respectively, and $\epsilon$ is a small constant for numerical stability. Fig.~\ref{fig8} shows the distributions of the normalised contributions by the tabular and network data.

It can be observed that both modalities contribute meaningfully to the final fused representation with the distributions centred around the midpoint of the normalised scale. The mean contribution from the network modality is slightly higher than that of the tabular modality, suggesting that, in the current task, the GNN embeddings exert a stronger effect in shaping the final representation. This outcome aligns with the intuition that structural information encoded in the network topology can enhance learning, especially when node interactions are important for prediction. However, the overlap between the two distributions indicates that the model benefits from both sources of data and does not ignore either modality, highlighting the advantage of modelling interactions between the two modalities rather than relying on either in isolation.

\subsection{Effects of edge direction and weights}
\label{Effects of edge direction and weights}
Table~\ref{tab5} explores the impact of including edge directionality and weights on the performance of the hybrid Concat-Att-GAT model for double-layer networks. In the financial transaction layer, the edge direction represents the flow of transactions between two firms, and the weight is determined by dividing the total sum of the transaction values by the number of transactions over a period of six months, following the approaches commonly used in transaction network studies \citep{saxena2021banking, elliott2019anomaly}. A logarithmic transformation is applied to the edge weights to enhance training stability and mitigate the risk of gradient explosion that can arise from excessively large weight values \citep{li2024graph}. For the common ownership layer, the edges are bidirectional, and their weights are uniformly set at 1.

\vspace{+6pt}
\begin{table}[hbt!]
\scriptsize
\centering
\caption{Performance of the best-performing proposed model on directed and/or weighted networks.}
\begin{tabular}{lc|cc|cc}
\toprule
\multicolumn{2}{c|}{Model}& \multicolumn{2}{c|}{Edges}& \multicolumn{2}{c}{Double Layer: FT-CO}\\
Strategy&\multicolumn{1}{c|}{GNN}&Directed&\multicolumn{1}{c|}{Weighted} & \multicolumn{1}{c}{AUC} & \multicolumn{1}{c}{AUCPR}\\
\midrule
\multirow{3}{*}{Hybrid Concat-Att}&{}&$\checkmark$&$\times$ & $0.872\pm0.010$ & $0.407\pm0.009$\\
&GAT&$\times$&$\checkmark$ & $0.878\pm0.012$ & $0.412\pm0.010$\\
&{}&$\checkmark$&$\checkmark$ & $\mathbf{0.879\pm0.011}$ & $\mathbf{0.414\pm0.010}$\\
\bottomrule
\end{tabular}
\label{tab5}
\end{table}

The results suggest that incorporating edge direction and weights improves the performance somewhat, increasing AUC to 0.879 and AUCPR to 0.414, although the confidence intervals of the three model variations overlap. This modest improvement may be attributed to the added granularity and realism provided by these characteristics. The directed edges capture the asymmetry in financial relationships, which can affect the flow of credit risk contagion. Edge weighting, which appears to provide richer information than edge orientation, reflects the varying intensities of the connections between SMEs, allowing the model to differentiate between stronger and weaker financial ties. These enhancements provide deeper insight into the mechanisms of credit risk propagation, leading to more accurate predictions.

\begin{figure}[hbt!]
\includegraphics[scale=0.18]{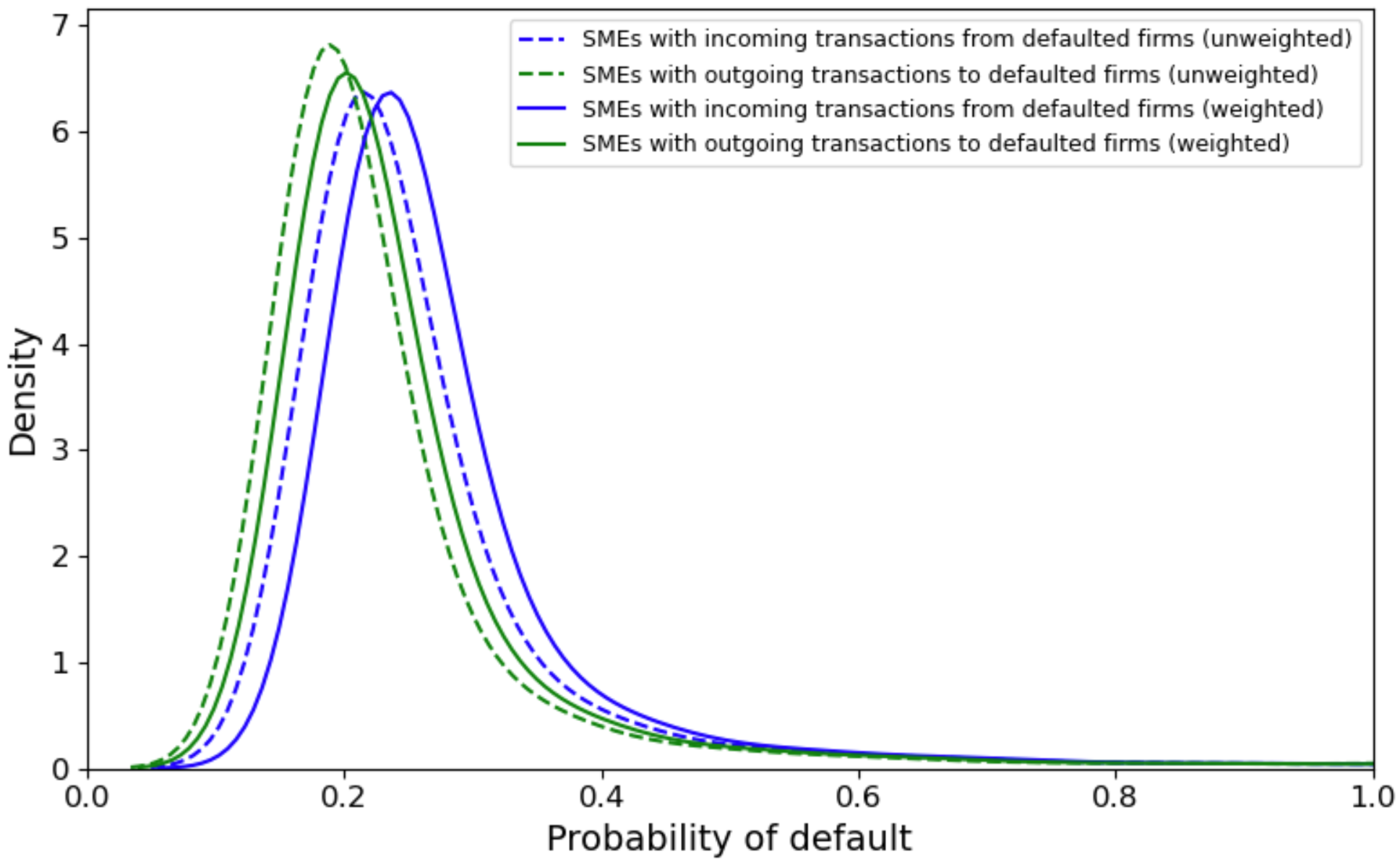}
\centering
\caption{Kernel density estimates of the estimated probabilities of default for SMEs exposed to defaulters via incoming or outgoing financial transactions, in the weighted and unweighted networks.}
\label{fig9}
\end{figure}

An interesting question about directional network models is whether they find that the default risk of a firm is more adversely affected by default events experienced by its payees (which may, in the context of a supply chain, be some of the firm's direct suppliers) or its payers (i.e.\@ by the customers of the firm). In the transaction graph layer, this corresponds to a node being connected to other defaulted nodes via outgoing or incoming edges, respectively. To shed light on this, Fig.~\ref{fig9} contains kernel density estimates of the estimated probabilities of default for SMEs that are linked to defaulters via outgoing or incoming transactions. Separate distributions are included for the weighted and unweighted networks.

The four density distributions shown in Fig.~\ref{fig9} appear to be different. Notably, SMEs that have been at the receiving end of transactions from a defaulted firm tend to be assigned a higher probability of default than those that have initiated transactions towards defaulted firms. This pattern is observed in both the weighted and unweighted networks, although the effect is more pronounced in the weighted case, where stronger financial ties can amplify risk transmission.

\subsection{Interpretability of the architecture}
\label{Interpretability of the architecture}
\begin{figure}[hbt!]
\centering
\begin{subfigure}[b]{0.55\textwidth}
\includegraphics[scale=0.165]{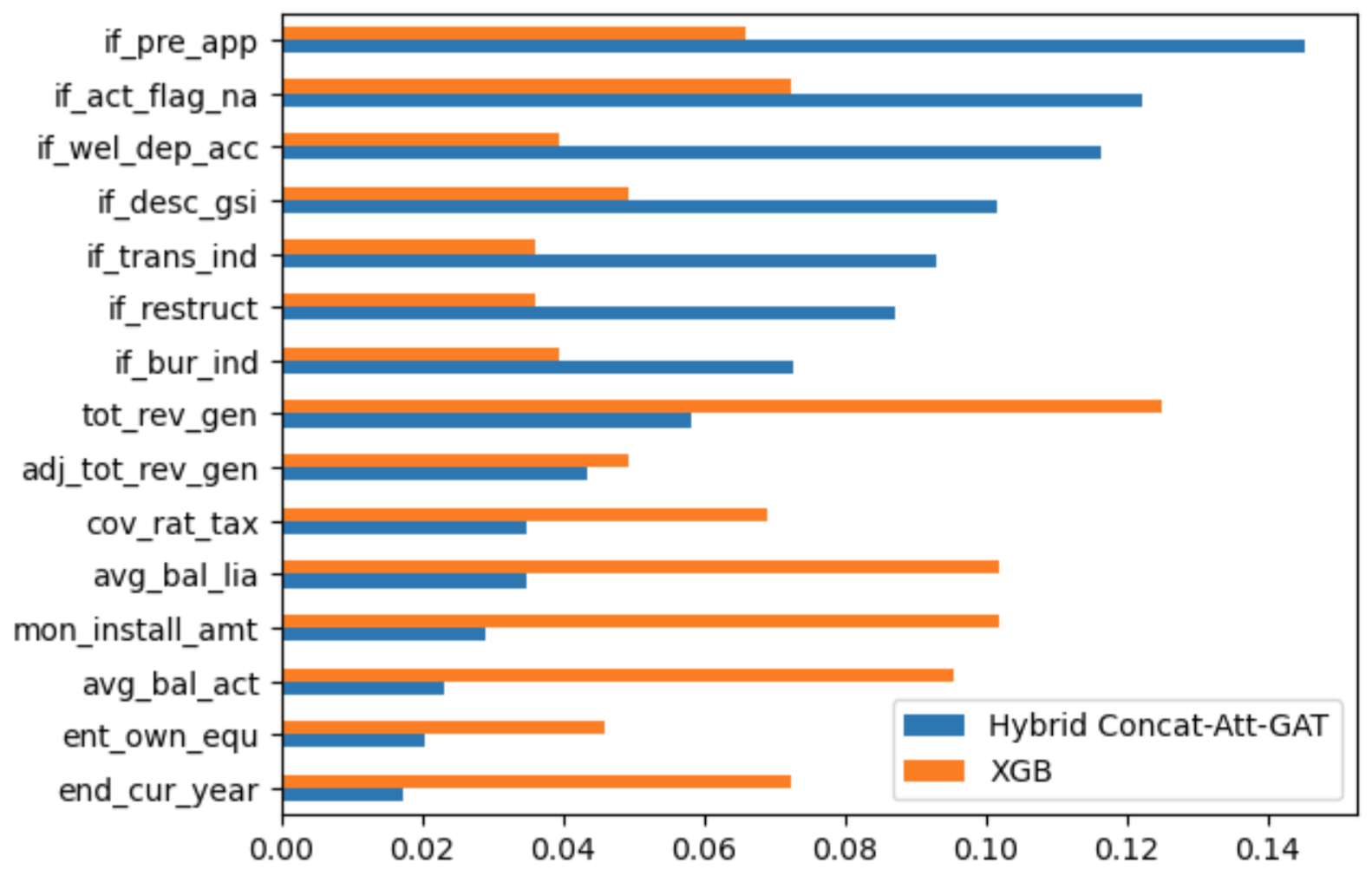}
\caption{Relative importance for Hybrid Concat-Att-GAT and XGB.}
\label{fig10a}
\end{subfigure}
\quad
\begin{subfigure}[b]{0.42\textwidth}
\includegraphics[scale=0.35]{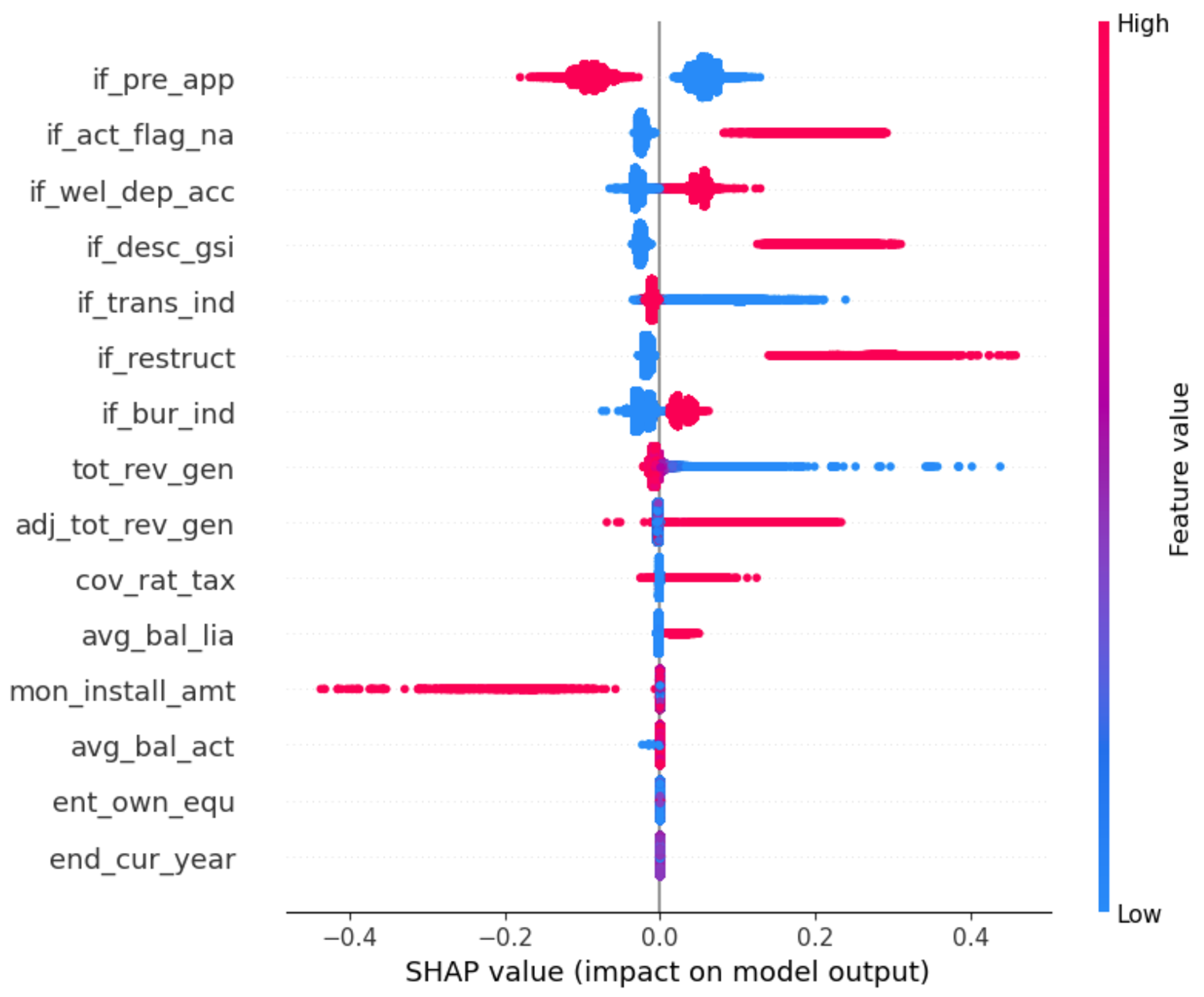}
\caption{Shapley values for Hybrid Concat-Att-GAT.}
\label{fig10b}
\end{subfigure}
\caption{Summary of feature importance.}
\label{fig10}
\end{figure}

Since the hybrid Concat-Att-GAT model applied to directed double-layer networks yielded the best predictive performance in the previous subsection, we now employ the Shapley approach \citep{lundberg2017unified} to further analyse this model. This method allows us to assess the relative importance of each node feature and quantify its contribution to the model outputs. Fig~\ref{fig10a} provides a ranking of the most important node features in the aforementioned model, along with their importance in the best baseline model, i.e.\@ XGB. In addition to this, Fig.\ref{fig10b} depicts how these features affect the predictions of the proposed model. Here, higher SHAP values for a (colour-coded) feature value push the model towards predicting $1$ (default); lower SHAP values drive it towards $0$ (non-default).

As seen from Fig.~\ref{fig10a}, the most important features in the hybrid Concat-Att-GAT model include several internal or external status indicators, such as `if\textunderscore pre\textunderscore app' (pre-approval status), `if\textunderscore act\textunderscore flag\textunderscore na' (missing active flag indicator), `if\textunderscore trans\textunderscore ind' (transaction indicator assigned), `if\textunderscore restruct' (restructuring indicator), and `if\textunderscore bur\textunderscore ind' (bureau indicator assigned). In contrast, XGB places greater importance on financial metrics, including income statement items such as `tot\textunderscore rev\textunderscore gen' (total revenue generated) and balance sheet measures such as `avg\textunderscore bal\textunderscore act' (average active assets) and `avg\textunderscore bal\textunderscore lia' (average liabilities), suggesting a preference for numerical features.

Fig.~\ref{fig10b} shows that the effects of these features generally align with expectations. For example, for the top variable, `if\textunderscore pre\textunderscore app', a positive pre-approval status (coded in red) tends to reduce the predicted default risk for the company, whereas negative values (coded in blue) increase it. Missing information (e.g.\@ positive values for `if\textunderscore act\textunderscore flag\textunderscore na') tends to increase risk. So does, for example, any indication that the firm has undergone restructuring (see `if\textunderscore restruct'), which is again an intuitive finding. 
Low values (depicted as blue dots) for `tot\textunderscore rev\textunderscore gen' are strongly associated with elevated default risk (in other words, firms that generate little revenue are deemed higher risk). Having greater liabilities is also seen to add to the risk, albeit more modestly (see, for example, `avg\textunderscore bal\textunderscore lia'). Being deemed lower risk, higher monthly instalments might be linked to larger firms that are capable of handling larger loans or credit obligations. Some of the other financial metrics, as was also apparent from Fig.~\ref{fig10a}, appear to be less influential in the model.

\section{Discussion}
\label{Discussion}

This study sought to address three key research questions: 1) whether incorporating explicit network data into predictive models improves their accuracy or provides a deeper understanding of how credit risk spreads among interconnected SMEs; and if this is the case, determine the optimal method for integrating network data with structured data; 2) whether multilayer networks offer advantages over single-layer networks; and if so, identify the type of connection that is more informative; and 3) whether additional insights can be gained by considering the directionality and intensity of these connections. 

Our findings indeed indicate that the integration of network data with traditional structured data improves the performance of the model. The results suggest that the bimodal model employing a hybrid fusion strategy with a cross-attention approach is the most effective. The use of multilayer networks instead of single-layer networks is shown to be more informative and effective in capturing the complexities of borrower interactions and credit risk contagion, which is consistent with other recent studies in the credit risk area \citep{oskarsdottir2021multilayer, zandi2024attention}. Furthermore, considering the directionality and intensity of these connections provides added insight into the propagation of risk, leading to more accurate predictions. This could be due to supply chain dependencies and financial contagion. In supply chains, upstream supplier distress can disrupt production and increase costs for downstream firms \citep{agca2022credit, tabachova2024estimating}. However, our results indicate that buyer-side defaults have an even greater impact on SMEs, as lost revenues and unpaid receivables may directly strain a firm's liquidity. This aligns with evidence that firms with concentrated client bases (as is common for many SMEs) face higher debt costs and risk exposure \citep{dhaliwal2016customer}, and that their use of trade credit can amplify the transmission of financial shocks through supply chains \citep{agca2022credit}.

The findings of this research have significant societal implications, as improving financial inclusion can support the growth of SMEs and broader economic development \citep{boachie2024effect}. Even in advanced economies, obtaining credit remains a challenging task for many SMEs,  especially for recently established firms without extensive financial histories  \citep{lee2015access}. Our approach allows financial institutions to assess credit risk more accurately, thus lowering lending costs and, in turn, enabling greater credit availability for SMEs. In emerging economies, where traditional financial records are often sparse, there are distinct opportunities to apply similar techniques. Many SMEs in these regions operate in environments where formal banking infrastructure is limited, yet engage in extensive interactions and are highly interdependent \citep{wang2016biggest}. These interdependencies can be derived from various sources such as call detail records, mobile money services, social media activities, community contributions, or geolocation data, among others. Our approach would enable the use of such interaction data to provide an alternative means of credit assessment, thus facilitating access to finance for a broader range of businesses. As for the lender, the ability to model credit risk contagion within SME networks can improve the resilience of a lender’s portfolio \citep{anagnostou2018incorporating}. By identifying how financial distress propagates through observed inter-firm connections, financial institutions can better anticipate correlated defaults within their client base and adjust risk management strategies accordingly. While our dataset only captures the slice of the market observed by one lender, these insights can nonetheless inform portfolio-level decision making and contribute to more prudent credit allocation.

\section{Conclusions}
\label{Conclusions}

This study demonstrated the value of incorporating multilayer network data and advanced deep learning techniques in predicting SME default risk. By integrating network-based information with traditional structured data, we showed that predictive models can more effectively assess the default risk of SMEs. To evaluate our approach, we used a large dataset from a major financial institution, deriving both single- and double-layer networks from company ownership structures and individual inter-firm financial transactions observed by the lender. In so doing, we were able to capture explicit supply chain relations between firms. We experimented with unimodal and bimodal architectures, employing two types of GNN, and testing various fusion strategies for bimodal models. Our results showed that bimodal learning, particularly when applying a hybrid fusion strategy and cross-attention, enhances predictive accuracy and provides deeper insight into risk propagation. We observed that the network modality has a slightly greater influence on the fused representation than the tabular modality. The results also showed that multilayer networks are able to extract more information from the data, leading to more accurate risk assessments. We found that employing networks with directed and weighted edges further boosts predictive performance, confirming that the strength and intensity of inter-firm links affects credit risk propagation. Lastly, we showed how Shapley analysis can shed light on the relative importance of different features and how this ranking may change as a result of adding the network modality.

Future research could explore several avenues to build on these findings. First, expanding the scope of network data to include additional types of interactions such as social media connections or geographic proximity could provide a more holistic view of SME networks. Social media data, for example, could offer insights into potential inter-firm linkages --- such as business partnerships, customer-supplier mentions, or co-membership in professional networks. Geographic proximity data could help model localised economic conditions and their impact on SMEs \citep{calabrese2019birds}. Second, advances in fusion techniques may further improve predictive performance, revealing more effective ways to integrate multimodal data. Lastly, applying the proposed methods in different contexts would test their robustness and adaptability. Lenders can use the framework to build models tailored to their own data, and testing across diverse scenarios would help ensure the methods’ reliability.

\section*{Acknowledgements}
\label{Acknowledgements}

The first author acknowledges the support of the Natural Sciences and Engineering Research Council (NSERC) of Canada through the Canada Graduate Scholarships – Doctoral (CGS D) program. The second and fifth authors acknowledge the support of the Economic and Social Research Council (ESRC) [grant number ES/P000673/1]. The fourth author acknowledges the support of the Icelandic Research Fund (IRF) [grant number 228511-051]. The last author acknowledges the support of the NSERC [discovery grant RGPIN-2020-07114]. This research was undertaken, in part, thanks to funding from the Canada Research Chairs program [CRC-2018-00082]. The authors acknowledge the use of the IRIDIS High Performance Computing Facility and associated support services at the University of Southampton in the completion of this work.

\bibliographystyle{model5-names.bst}
{\bibliography{References.bib}}

\begin{thebibliography}{71}
\expandafter\ifx\csname natexlab\endcsname\relax\def\natexlab#1{#1}\fi
\providecommand{\url}[1]{\texttt{#1}}
\providecommand{\href}[2]{#2}
\providecommand{\path}[1]{#1}
\providecommand{\DOIprefix}{doi:}
\providecommand{\ArXivprefix}{arXiv:}
\providecommand{\URLprefix}{URL: }
\providecommand{\Pubmedprefix}{pmid:}
\providecommand{\doi}[1]{\href{http://dx.doi.org/#1}{\path{#1}}}
\providecommand{\Pubmed}[1]{\href{pmid:#1}{\path{#1}}}
\providecommand{\bibinfo}[2]{#2}
\ifx\xfnm\relax \def\xfnm[#1]{\unskip,\space#1}\fi
%Type = Article
\bibitem[{Agca et~al.(2022)Agca, Babich, Birge \& Wu}]{agca2022credit}
\bibinfo{author}{Agca, S.}, \bibinfo{author}{Babich, V.}, \bibinfo{author}{Birge, J.~R.}, \& \bibinfo{author}{Wu, J.} (\bibinfo{year}{2022}).
\newblock \bibinfo{title}{Credit shock propagation along supply chains: Evidence from the {CDS} market}.
\newblock {\it \bibinfo{journal}{Management Science}\/},  {\it \bibinfo{volume}{68}\/}, \bibinfo{pages}{6506--6538}.
%Type = Article
\bibitem[{Allen \& Gale(2000)}]{allen2000financial}
\bibinfo{author}{Allen, F.}, \& \bibinfo{author}{Gale, D.} (\bibinfo{year}{2000}).
\newblock \bibinfo{title}{Financial contagion}.
\newblock {\it \bibinfo{journal}{Journal of Political Economy}\/},  {\it \bibinfo{volume}{108}\/}, \bibinfo{pages}{1--33}.
%Type = Article
\bibitem[{Altman(1968)}]{altman1968financial}
\bibinfo{author}{Altman, E.~I.} (\bibinfo{year}{1968}).
\newblock \bibinfo{title}{Financial ratios, discriminant analysis and the prediction of corporate bankruptcy}.
\newblock {\it \bibinfo{journal}{The Journal of Finance}\/},  {\it \bibinfo{volume}{23}\/}, \bibinfo{pages}{589--609}.
%Type = Article
\bibitem[{Anagnostou et~al.(2018)Anagnostou, Sourabh \& Kandhai}]{anagnostou2018incorporating}
\bibinfo{author}{Anagnostou, I.}, \bibinfo{author}{Sourabh, S.}, \& \bibinfo{author}{Kandhai, D.} (\bibinfo{year}{2018}).
\newblock \bibinfo{title}{Incorporating contagion in portfolio credit risk models using network theory}.
\newblock {\it \bibinfo{journal}{Complexity}\/},  {\it \bibinfo{volume}{2018}\/}, \bibinfo{pages}{6076173}.
%Type = Article
\bibitem[{Bakhtiari et~al.(2020)Bakhtiari, Breunig, Magnani \& Zhang}]{bakhtiari2020financial}
\bibinfo{author}{Bakhtiari, S.}, \bibinfo{author}{Breunig, R.}, \bibinfo{author}{Magnani, L.}, \& \bibinfo{author}{Zhang, J.} (\bibinfo{year}{2020}).
\newblock \bibinfo{title}{Financial constraints and small and medium enterprises: A review}.
\newblock {\it \bibinfo{journal}{Economic Record}\/},  {\it \bibinfo{volume}{96}\/}, \bibinfo{pages}{506--523}.
%Type = Article
\bibitem[{Baltru{\v{s}}aitis et~al.(2018)Baltru{\v{s}}aitis, Ahuja \& Morency}]{baltruvsaitis2018multimodal}
\bibinfo{author}{Baltru{\v{s}}aitis, T.}, \bibinfo{author}{Ahuja, C.}, \& \bibinfo{author}{Morency, L.-P.} (\bibinfo{year}{2018}).
\newblock \bibinfo{title}{Multimodal machine learning: A survey and taxonomy}.
\newblock {\it \bibinfo{journal}{IEEE Transactions on Pattern Analysis and Machine Intelligence}\/},  {\it \bibinfo{volume}{41}\/}, \bibinfo{pages}{423--443}.
%Type = Book
\bibitem[{Barabási \& Pósfai(2016)}]{barabasi2016network}
\bibinfo{author}{Barabási, A.-L.}, \& \bibinfo{author}{Pósfai, M.} (\bibinfo{year}{2016}).
\newblock {\it \bibinfo{title}{Network science}\/}.
\newblock \bibinfo{publisher}{Cambridge University Press}.
%Type = Article
\bibitem[{Beaver et~al.(2019)Beaver, Cascino, Correia \& McNichols}]{beaver2019group}
\bibinfo{author}{Beaver, W.~H.}, \bibinfo{author}{Cascino, S.}, \bibinfo{author}{Correia, M.}, \& \bibinfo{author}{McNichols, M.~F.} (\bibinfo{year}{2019}).
\newblock \bibinfo{title}{Group affiliation and default prediction}.
\newblock {\it \bibinfo{journal}{Management Science}\/},  {\it \bibinfo{volume}{65}\/}, \bibinfo{pages}{3559--3584}.
%Type = Article
\bibitem[{Berloco et~al.(2021)Berloco, De~Francisci~Morales, Frassineti, Greco, Kumarasinghe, Lamieri, Massaro, Miola \& Yang}]{berloco2021predicting}
\bibinfo{author}{Berloco, C.}, \bibinfo{author}{De~Francisci~Morales, G.}, \bibinfo{author}{Frassineti, D.}, \bibinfo{author}{Greco, G.}, \bibinfo{author}{Kumarasinghe, H.}, \bibinfo{author}{Lamieri, M.}, \bibinfo{author}{Massaro, E.}, \bibinfo{author}{Miola, A.}, \& \bibinfo{author}{Yang, S.} (\bibinfo{year}{2021}).
\newblock \bibinfo{title}{Predicting corporate credit risk: Network contagion via trade credit}.
\newblock {\it \bibinfo{journal}{PLoS One}\/},  {\it \bibinfo{volume}{16}\/}, \bibinfo{pages}{e0250115}.
%Type = Article
\bibitem[{Boachie \& Adu-Darko(2024)}]{boachie2024effect}
\bibinfo{author}{Boachie, C.}, \& \bibinfo{author}{Adu-Darko, E.} (\bibinfo{year}{2024}).
\newblock \bibinfo{title}{The effect of financial inclusion on economic growth: The role of human capital development}.
\newblock {\it \bibinfo{journal}{Cogent Social Sciences}\/},  {\it \bibinfo{volume}{10}\/}, \bibinfo{pages}{2346118}.
%Type = Article
\bibitem[{Borgatti \& Halgin(2011)}]{borgatti2011network}
\bibinfo{author}{Borgatti, S.~P.}, \& \bibinfo{author}{Halgin, D.~S.} (\bibinfo{year}{2011}).
\newblock \bibinfo{title}{On network theory}.
\newblock {\it \bibinfo{journal}{Organization Science}\/},  {\it \bibinfo{volume}{22}\/}, \bibinfo{pages}{1168--1181}.
%Type = Article
\bibitem[{Borisov et~al.(2022)Borisov, Leemann, Se{\ss}ler, Haug, Pawelczyk \& Kasneci}]{borisov2022deep}
\bibinfo{author}{Borisov, V.}, \bibinfo{author}{Leemann, T.}, \bibinfo{author}{Se{\ss}ler, K.}, \bibinfo{author}{Haug, J.}, \bibinfo{author}{Pawelczyk, M.}, \& \bibinfo{author}{Kasneci, G.} (\bibinfo{year}{2022}).
\newblock \bibinfo{title}{Deep neural networks and tabular data: A survey}.
\newblock {\it \bibinfo{journal}{IEEE Transactions on Neural Networks and Learning Systems}\/},  {\it \bibinfo{volume}{35}\/}, \bibinfo{pages}{7499--7519}.
%Type = Article
\bibitem[{Boulahia et~al.(2021)Boulahia, Amamra, Madi \& Daikh}]{boulahia2021early}
\bibinfo{author}{Boulahia, S.~Y.}, \bibinfo{author}{Amamra, A.}, \bibinfo{author}{Madi, M.~R.}, \& \bibinfo{author}{Daikh, S.} (\bibinfo{year}{2021}).
\newblock \bibinfo{title}{Early, intermediate and late fusion strategies for robust deep learning-based multimodal action recognition}.
\newblock {\it \bibinfo{journal}{Machine Vision and Applications}\/},  {\it \bibinfo{volume}{32}\/}, \bibinfo{pages}{121}.
%Type = Article
\bibitem[{Bravo et~al.(2013)Bravo, Maldonado \& Weber}]{bravo2013granting}
\bibinfo{author}{Bravo, C.}, \bibinfo{author}{Maldonado, S.}, \& \bibinfo{author}{Weber, R.} (\bibinfo{year}{2013}).
\newblock \bibinfo{title}{Granting and managing loans for micro-entrepreneurs: New developments and practical experiences}.
\newblock {\it \bibinfo{journal}{European Journal of Operational Research}\/},  {\it \bibinfo{volume}{227}\/}, \bibinfo{pages}{358--366}.
%Type = Inproceedings
\bibitem[{Bravo \& {\'O}skarsd{\'o}ttir(2020)}]{bravo2020evolution}
\bibinfo{author}{Bravo, C.}, \& \bibinfo{author}{{\'O}skarsd{\'o}ttir, M.} (\bibinfo{year}{2020}).
\newblock \bibinfo{title}{Evolution of credit risk using a personalized pagerank algorithm for multilayer networks}.
\newblock In {\it \bibinfo{booktitle}{KDD MLF 2020: KDD Workshop on Machine Learning in Finance}\/}.
%Type = Article
\bibitem[{Breiman(2001)}]{breiman2001random}
\bibinfo{author}{Breiman, L.} (\bibinfo{year}{2001}).
\newblock \bibinfo{title}{Random forests}.
\newblock {\it \bibinfo{journal}{Machine Learning}\/},  {\it \bibinfo{volume}{45}\/}, \bibinfo{pages}{5--32}.
%Type = Article
\bibitem[{Calabrese et~al.(2019)Calabrese, Andreeva \& Ansell}]{calabrese2019birds}
\bibinfo{author}{Calabrese, R.}, \bibinfo{author}{Andreeva, G.}, \& \bibinfo{author}{Ansell, J.} (\bibinfo{year}{2019}).
\newblock \bibinfo{title}{“{B}irds of a feather” fail together: Exploring the nature of dependency in {SME} defaults}.
\newblock {\it \bibinfo{journal}{Risk Analysis}\/},  {\it \bibinfo{volume}{39}\/}, \bibinfo{pages}{71--84}.
%Type = Article
\bibitem[{Chango et~al.(2022)Chango, Lara, Cerezo \& Romero}]{chango2022review}
\bibinfo{author}{Chango, W.}, \bibinfo{author}{Lara, J.~A.}, \bibinfo{author}{Cerezo, R.}, \& \bibinfo{author}{Romero, C.} (\bibinfo{year}{2022}).
\newblock \bibinfo{title}{A review on data fusion in multimodal learning analytics and educational data mining}.
\newblock {\it \bibinfo{journal}{Wiley Interdisciplinary Reviews: Data Mining and Knowledge Discovery}\/},  {\it \bibinfo{volume}{12}\/}, \bibinfo{pages}{e1458}.
%Type = Inproceedings
\bibitem[{Chen \& Guestrin(2016)}]{chen2016xgboost}
\bibinfo{author}{Chen, T.}, \& \bibinfo{author}{Guestrin, C.} (\bibinfo{year}{2016}).
\newblock \bibinfo{title}{{XGB}oost: A scalable tree boosting system}.
\newblock In {\it \bibinfo{booktitle}{Proceedings of the 22nd ACM SIGKDD International Conference on Knowledge Discovery and Data Mining}\/} (pp. \bibinfo{pages}{785--794}).
%Type = Article
\bibitem[{De~Bock et~al.(2024)De~Bock, Coussement, De~Caigny, Slowi{\'n}ski, Baesens, Boute, Choi, Delen, Kraus, Lessmann et~al.}]{de2023explainable}
\bibinfo{author}{De~Bock, K.~W.}, \bibinfo{author}{Coussement, K.}, \bibinfo{author}{De~Caigny, A.}, \bibinfo{author}{Slowi{\'n}ski, R.}, \bibinfo{author}{Baesens, B.}, \bibinfo{author}{Boute, R.~N.}, \bibinfo{author}{Choi, T.-M.}, \bibinfo{author}{Delen, D.}, \bibinfo{author}{Kraus, M.}, \bibinfo{author}{Lessmann, S.} et~al. (\bibinfo{year}{2024}).
\newblock \bibinfo{title}{Explainable {AI} for operational research: A defining framework, methods, applications, and a research agenda}.
\newblock {\it \bibinfo{journal}{European Journal of Operational Research}\/},  {\it \bibinfo{volume}{317}\/}, \bibinfo{pages}{249--272}.
%Type = Article
\bibitem[{Dhaliwal et~al.(2016)Dhaliwal, Judd, Serfling \& Shaikh}]{dhaliwal2016customer}
\bibinfo{author}{Dhaliwal, D.}, \bibinfo{author}{Judd, J.~S.}, \bibinfo{author}{Serfling, M.}, \& \bibinfo{author}{Shaikh, S.} (\bibinfo{year}{2016}).
\newblock \bibinfo{title}{Customer concentration risk and the cost of equity capital}.
\newblock {\it \bibinfo{journal}{Journal of Accounting and Economics}\/},  {\it \bibinfo{volume}{61}\/}, \bibinfo{pages}{23--48}.
%Type = Article
\bibitem[{Elliott et~al.(2019)Elliott, Cucuringu, Luaces, Reidy \& Reinert}]{elliott2019anomaly}
\bibinfo{author}{Elliott, A.}, \bibinfo{author}{Cucuringu, M.}, \bibinfo{author}{Luaces, M.~M.}, \bibinfo{author}{Reidy, P.}, \& \bibinfo{author}{Reinert, G.} (\bibinfo{year}{2019}).
\newblock \bibinfo{title}{Anomaly detection in networks with application to financial transaction networks}.
\newblock {\it \bibinfo{journal}{arXiv preprint arXiv:1901.00402}\/}, .
%Type = Article
\bibitem[{Fenech et~al.(2015)Fenech, Vosgha \& Shafik}]{fenech2015loan}
\bibinfo{author}{Fenech, J.~P.}, \bibinfo{author}{Vosgha, H.}, \& \bibinfo{author}{Shafik, S.} (\bibinfo{year}{2015}).
\newblock \bibinfo{title}{Loan default correlation using an {A}rchimedean copula approach: A case for recalibration}.
\newblock {\it \bibinfo{journal}{Economic Modelling}\/},  {\it \bibinfo{volume}{47}\/}, \bibinfo{pages}{340--354}.
%Type = Article
\bibitem[{Giesecke \& Weber(2004)}]{giesecke2004cyclical}
\bibinfo{author}{Giesecke, K.}, \& \bibinfo{author}{Weber, S.} (\bibinfo{year}{2004}).
\newblock \bibinfo{title}{Cyclical correlations, credit contagion, and portfolio losses}.
\newblock {\it \bibinfo{journal}{Journal of Banking \& Finance}\/},  {\it \bibinfo{volume}{28}\/}, \bibinfo{pages}{3009--3036}.
%Type = Article
\bibitem[{Gneiting \& Raftery(2007)}]{gneiting2007strictly}
\bibinfo{author}{Gneiting, T.}, \& \bibinfo{author}{Raftery, A.~E.} (\bibinfo{year}{2007}).
\newblock \bibinfo{title}{Strictly proper scoring rules, prediction, and estimation}.
\newblock {\it \bibinfo{journal}{Journal of the American Statistical Association}\/},  {\it \bibinfo{volume}{102}\/}, \bibinfo{pages}{359--378}.
%Type = Article
\bibitem[{Gunnarsson et~al.(2021)Gunnarsson, Vanden~Broucke, Baesens, {\'O}skarsd{\'o}ttir \& Lemahieu}]{gunnarsson2021deep}
\bibinfo{author}{Gunnarsson, B.~R.}, \bibinfo{author}{Vanden~Broucke, S.}, \bibinfo{author}{Baesens, B.}, \bibinfo{author}{{\'O}skarsd{\'o}ttir, M.}, \& \bibinfo{author}{Lemahieu, W.} (\bibinfo{year}{2021}).
\newblock \bibinfo{title}{Deep learning for credit scoring: Do or don’t?}
\newblock {\it \bibinfo{journal}{European Journal of Operational Research}\/},  {\it \bibinfo{volume}{295}\/}, \bibinfo{pages}{292--305}.
%Type = Book
\bibitem[{Hosmer~Jr. et~al.(2013)Hosmer~Jr., Lemeshow \& Sturdivant}]{hosmer2013applied}
\bibinfo{author}{Hosmer~Jr., D.~W.}, \bibinfo{author}{Lemeshow, S.}, \& \bibinfo{author}{Sturdivant, R.~X.} (\bibinfo{year}{2013}).
\newblock {\it \bibinfo{title}{Applied logistic regression}\/}.
\newblock \bibinfo{publisher}{John Wiley \& Sons}.
%Type = Article
\bibitem[{Huang \& Yang(2024)}]{huang2024time}
\bibinfo{author}{Huang, C.}, \& \bibinfo{author}{Yang, Y.} (\bibinfo{year}{2024}).
\newblock \bibinfo{title}{Time series feature redundancy paradox: An empirical study based on mortgage default prediction}.
\newblock {\it \bibinfo{journal}{arXiv preprint arXiv:2501.00034}\/}, .
%Type = Article
\bibitem[{Iori et~al.(2008)Iori, De~Masi, Precup, Gabbi \& Caldarelli}]{iori2008network}
\bibinfo{author}{Iori, G.}, \bibinfo{author}{De~Masi, G.}, \bibinfo{author}{Precup, O.~V.}, \bibinfo{author}{Gabbi, G.}, \& \bibinfo{author}{Caldarelli, G.} (\bibinfo{year}{2008}).
\newblock \bibinfo{title}{A network analysis of the {I}talian overnight money market}.
\newblock {\it \bibinfo{journal}{Journal of Economic Dynamics and Control}\/},  {\it \bibinfo{volume}{32}\/}, \bibinfo{pages}{259--278}.
%Type = Article
\bibitem[{Jackson \& Pernoud(2021)}]{jackson2021systemic}
\bibinfo{author}{Jackson, M.~O.}, \& \bibinfo{author}{Pernoud, A.} (\bibinfo{year}{2021}).
\newblock \bibinfo{title}{Systemic risk in financial networks: A survey}.
\newblock {\it \bibinfo{journal}{Annual Review of Economics}\/},  {\it \bibinfo{volume}{13}\/}, \bibinfo{pages}{171--202}.
%Type = Article
\bibitem[{Kivel{\"a} et~al.(2014)Kivel{\"a}, Arenas, Barthelemy, Gleeson, Moreno \& Porter}]{kivela2014multilayer}
\bibinfo{author}{Kivel{\"a}, M.}, \bibinfo{author}{Arenas, A.}, \bibinfo{author}{Barthelemy, M.}, \bibinfo{author}{Gleeson, J.~P.}, \bibinfo{author}{Moreno, Y.}, \& \bibinfo{author}{Porter, M.~A.} (\bibinfo{year}{2014}).
\newblock \bibinfo{title}{Multilayer networks}.
\newblock {\it \bibinfo{journal}{Journal of Complex Networks}\/},  {\it \bibinfo{volume}{2}\/}, \bibinfo{pages}{203--271}.
%Type = Article
\bibitem[{Korangi et~al.(2023)Korangi, Mues \& Bravo}]{korangi2023transformer}
\bibinfo{author}{Korangi, K.}, \bibinfo{author}{Mues, C.}, \& \bibinfo{author}{Bravo, C.} (\bibinfo{year}{2023}).
\newblock \bibinfo{title}{A transformer-based model for default prediction in mid-cap corporate markets}.
\newblock {\it \bibinfo{journal}{European Journal of Operational Research}\/},  {\it \bibinfo{volume}{308}\/}, \bibinfo{pages}{306--320}.
%Type = Article
\bibitem[{LeCun et~al.(2015)LeCun, Bengio \& Hinton}]{lecun2015deep}
\bibinfo{author}{LeCun, Y.}, \bibinfo{author}{Bengio, Y.}, \& \bibinfo{author}{Hinton, G.} (\bibinfo{year}{2015}).
\newblock \bibinfo{title}{Deep learning}.
\newblock {\it \bibinfo{journal}{Nature}\/},  {\it \bibinfo{volume}{521}\/}, \bibinfo{pages}{436--444}.
%Type = Inproceedings
\bibitem[{Lee et~al.(2018)Lee, Chen, Hua, Hu \& He}]{lee2018stacked}
\bibinfo{author}{Lee, K.-H.}, \bibinfo{author}{Chen, X.}, \bibinfo{author}{Hua, G.}, \bibinfo{author}{Hu, H.}, \& \bibinfo{author}{He, X.} (\bibinfo{year}{2018}).
\newblock \bibinfo{title}{Stacked cross attention for image-text matching}.
\newblock In {\it \bibinfo{booktitle}{Proceedings of the European Conference on Computer Vision (ECCV)}\/} (pp. \bibinfo{pages}{201--216}).
%Type = Article
\bibitem[{Lee et~al.(2015)Lee, Sameen \& Cowling}]{lee2015access}
\bibinfo{author}{Lee, N.}, \bibinfo{author}{Sameen, H.}, \& \bibinfo{author}{Cowling, M.} (\bibinfo{year}{2015}).
\newblock \bibinfo{title}{Access to finance for innovative {SME}s since the financial crisis}.
\newblock {\it \bibinfo{journal}{Research Policy}\/},  {\it \bibinfo{volume}{44}\/}, \bibinfo{pages}{370--380}.
%Type = Article
\bibitem[{Letizia \& Lillo(2019)}]{letizia2019corporate}
\bibinfo{author}{Letizia, E.}, \& \bibinfo{author}{Lillo, F.} (\bibinfo{year}{2019}).
\newblock \bibinfo{title}{Corporate payments networks and credit risk rating}.
\newblock {\it \bibinfo{journal}{EPJ Data Science}\/},  {\it \bibinfo{volume}{8}\/}, \bibinfo{pages}{21}.
%Type = Article
\bibitem[{Li et~al.(2024)Li, Shi \& van Leeuwen}]{li2024graph}
\bibinfo{author}{Li, Z.}, \bibinfo{author}{Shi, J.}, \& \bibinfo{author}{van Leeuwen, M.} (\bibinfo{year}{2024}).
\newblock \bibinfo{title}{Graph neural networks based log anomaly detection and explanation}.
\newblock {\it \bibinfo{journal}{arXiv preprint arXiv:2307.00527v3}\/}, .
%Type = Article
\bibitem[{Long et~al.(2022)Long, Jiang, Dimitrov \& Wang}]{long2022clues}
\bibinfo{author}{Long, J.}, \bibinfo{author}{Jiang, C.}, \bibinfo{author}{Dimitrov, S.}, \& \bibinfo{author}{Wang, Z.} (\bibinfo{year}{2022}).
\newblock \bibinfo{title}{Clues from networks: Quantifying relational risk for credit risk evaluation of {SME}s}.
\newblock {\it \bibinfo{journal}{Financial Innovation}\/},  {\it \bibinfo{volume}{8}\/}, \bibinfo{pages}{91}.
%Type = Article
\bibitem[{Lopez \& Saidenberg(2000)}]{lopez2000evaluating}
\bibinfo{author}{Lopez, J.~A.}, \& \bibinfo{author}{Saidenberg, M.~R.} (\bibinfo{year}{2000}).
\newblock \bibinfo{title}{Evaluating credit risk models}.
\newblock {\it \bibinfo{journal}{Journal of Banking \& Finance}\/},  {\it \bibinfo{volume}{24}\/}, \bibinfo{pages}{151--165}.
%Type = Article
\bibitem[{Lu et~al.(2025)Lu, Zhang, Su, Liu \& Yu}]{lu2025efficient}
\bibinfo{author}{Lu, S.}, \bibinfo{author}{Zhang, X.}, \bibinfo{author}{Su, Y.}, \bibinfo{author}{Liu, X.}, \& \bibinfo{author}{Yu, L.} (\bibinfo{year}{2025}).
\newblock \bibinfo{title}{Efficient multimodal learning for corporate credit risk prediction with an extended deep belief network}.
\newblock {\it \bibinfo{journal}{Annals of Operations Research}\/},  (pp. \bibinfo{pages}{1--38}).
%Type = Inproceedings
\bibitem[{Lundberg \& Lee(2017)}]{lundberg2017unified}
\bibinfo{author}{Lundberg, S.~M.}, \& \bibinfo{author}{Lee, S.-I.} (\bibinfo{year}{2017}).
\newblock \bibinfo{title}{A unified approach to interpreting model predictions}.
\newblock In {\it \bibinfo{booktitle}{Proceedings of the 31st International Conference on Neural Information Processing Systems (NIPS'17)}\/} (pp. \bibinfo{pages}{4768–--4777}).
%Type = Inproceedings
\bibitem[{Luo \& Kay(1988)}]{luo1988multisensor}
\bibinfo{author}{Luo, R.~C.}, \& \bibinfo{author}{Kay, M.~G.} (\bibinfo{year}{1988}).
\newblock \bibinfo{title}{Multisensor integration and fusion: Issues and approaches}.
\newblock In {\it \bibinfo{booktitle}{Sensor Fusion}\/} (pp. \bibinfo{pages}{42--49}).
\newblock volume \bibinfo{volume}{931}.
%Type = Article
\bibitem[{Mai et~al.(2019)Mai, Tian, Lee \& Ma}]{mai2019deep}
\bibinfo{author}{Mai, F.}, \bibinfo{author}{Tian, S.}, \bibinfo{author}{Lee, C.}, \& \bibinfo{author}{Ma, L.} (\bibinfo{year}{2019}).
\newblock \bibinfo{title}{Deep learning models for bankruptcy prediction using textual disclosures}.
\newblock {\it \bibinfo{journal}{European Journal of Operational Research}\/},  {\it \bibinfo{volume}{274}\/}, \bibinfo{pages}{743--758}.
%Type = Article
\bibitem[{Massa \& {\v{Z}}aldokas(2017)}]{massa2017information}
\bibinfo{author}{Massa, M.}, \& \bibinfo{author}{{\v{Z}}aldokas, A.} (\bibinfo{year}{2017}).
\newblock \bibinfo{title}{Information transfers among co-owned firms}.
\newblock {\it \bibinfo{journal}{Journal of Financial Intermediation}\/},  {\it \bibinfo{volume}{31}\/}, \bibinfo{pages}{77--92}.
%Type = Article
\bibitem[{Nagpal \& Bahar(2001)}]{nagpal2001measuring}
\bibinfo{author}{Nagpal, K.}, \& \bibinfo{author}{Bahar, R.} (\bibinfo{year}{2001}).
\newblock \bibinfo{title}{Measuring default correlation}.
\newblock {\it \bibinfo{journal}{Risk}\/},  {\it \bibinfo{volume}{14}\/}, \bibinfo{pages}{129--132}.
%Type = Article
\bibitem[{{\'O}skarsd{\'o}ttir \& Bravo(2021)}]{oskarsdottir2021multilayer}
\bibinfo{author}{{\'O}skarsd{\'o}ttir, M.}, \& \bibinfo{author}{Bravo, C.} (\bibinfo{year}{2021}).
\newblock \bibinfo{title}{Multilayer network analysis for improved credit risk prediction}.
\newblock {\it \bibinfo{journal}{Omega}\/},  {\it \bibinfo{volume}{105}\/}, \bibinfo{pages}{102520}.
%Type = Article
\bibitem[{{\'O}skarsd{\'o}ttir et~al.(2019){\'O}skarsd{\'o}ttir, Bravo, Sarraute, Vanthienen \& Baesens}]{oskarsdottir2019value}
\bibinfo{author}{{\'O}skarsd{\'o}ttir, M.}, \bibinfo{author}{Bravo, C.}, \bibinfo{author}{Sarraute, C.}, \bibinfo{author}{Vanthienen, J.}, \& \bibinfo{author}{Baesens, B.} (\bibinfo{year}{2019}).
\newblock \bibinfo{title}{The value of big data for credit scoring: Enhancing financial inclusion using mobile phone data and social network analytics}.
\newblock {\it \bibinfo{journal}{Applied Soft Computing}\/},  {\it \bibinfo{volume}{74}\/}, \bibinfo{pages}{26--39}.
%Type = Article
\bibitem[{Poria et~al.(2017)Poria, Cambria, Bajpai \& Hussain}]{poria2017review}
\bibinfo{author}{Poria, S.}, \bibinfo{author}{Cambria, E.}, \bibinfo{author}{Bajpai, R.}, \& \bibinfo{author}{Hussain, A.} (\bibinfo{year}{2017}).
\newblock \bibinfo{title}{A review of affective computing: From unimodal analysis to multimodal fusion}.
\newblock {\it \bibinfo{journal}{Information Fusion}\/},  {\it \bibinfo{volume}{37}\/}, \bibinfo{pages}{98--125}.
%Type = Article
\bibitem[{Ramachandram \& Taylor(2017)}]{ramachandram2017deep}
\bibinfo{author}{Ramachandram, D.}, \& \bibinfo{author}{Taylor, G.~W.} (\bibinfo{year}{2017}).
\newblock \bibinfo{title}{Deep multimodal learning: A survey on recent advances and trends}.
\newblock {\it \bibinfo{journal}{IEEE Signal Processing Magazine}\/},  {\it \bibinfo{volume}{34}\/}, \bibinfo{pages}{96--108}.
%Type = Article
\bibitem[{Rao et~al.(2023)Rao, Chatterjee, Nagaraju, Khan, Almusharraf \& Alharbi}]{rao2023fusion}
\bibinfo{author}{Rao, P.~K.}, \bibinfo{author}{Chatterjee, S.}, \bibinfo{author}{Nagaraju, K.}, \bibinfo{author}{Khan, S.~B.}, \bibinfo{author}{Almusharraf, A.}, \& \bibinfo{author}{Alharbi, A.~I.} (\bibinfo{year}{2023}).
\newblock \bibinfo{title}{Fusion of graph and tabular deep learning models for predicting chronic kidney disease}.
\newblock {\it \bibinfo{journal}{Diagnostics}\/},  {\it \bibinfo{volume}{13}\/}, \bibinfo{pages}{1981}.
%Type = Article
\bibitem[{Rishehchi~Fayyaz et~al.(2021)Rishehchi~Fayyaz, Rasouli \& Amiri}]{rishehchi2021data}
\bibinfo{author}{Rishehchi~Fayyaz, M.}, \bibinfo{author}{Rasouli, M.~R.}, \& \bibinfo{author}{Amiri, B.} (\bibinfo{year}{2021}).
\newblock \bibinfo{title}{A data-driven and network-aware approach for credit risk prediction in supply chain finance}.
\newblock {\it \bibinfo{journal}{Industrial Management \& Data Systems}\/},  {\it \bibinfo{volume}{121}\/}, \bibinfo{pages}{785--808}.
%Type = Inproceedings
\bibitem[{Saxena et~al.(2021)Saxena, Pei, Veldsink, van Ipenburg, Fletcher \& Pechenizkiy}]{saxena2021banking}
\bibinfo{author}{Saxena, A.}, \bibinfo{author}{Pei, Y.}, \bibinfo{author}{Veldsink, J.}, \bibinfo{author}{van Ipenburg, W.}, \bibinfo{author}{Fletcher, G.}, \& \bibinfo{author}{Pechenizkiy, M.} (\bibinfo{year}{2021}).
\newblock \bibinfo{title}{The banking transactions dataset and its comparative analysis with scale-free networks}.
\newblock In {\it \bibinfo{booktitle}{Proceedings of the 2021 IEEE/ACM International Conference on Advances in Social Networks Analysis and Mining}\/} (pp. \bibinfo{pages}{283--296}).
%Type = Article
\bibitem[{Smit \& Watkins(2012)}]{smit2012literature}
\bibinfo{author}{Smit, Y.}, \& \bibinfo{author}{Watkins, J.~A.} (\bibinfo{year}{2012}).
\newblock \bibinfo{title}{A literature review of small and medium enterprises ({SME}) risk management practices in {S}outh {A}frica}.
\newblock {\it \bibinfo{journal}{African Journal of Business Management}\/},  {\it \bibinfo{volume}{6}\/}, \bibinfo{pages}{6324--6330}.
%Type = Article
\bibitem[{Spatareanu et~al.(2023)Spatareanu, Manole, Kabiri \& Roland}]{spatareanu2023bank}
\bibinfo{author}{Spatareanu, M.}, \bibinfo{author}{Manole, V.}, \bibinfo{author}{Kabiri, A.}, \& \bibinfo{author}{Roland, I.} (\bibinfo{year}{2023}).
\newblock \bibinfo{title}{Bank default risk propagation along supply chains: Evidence from the {UK}}.
\newblock {\it \bibinfo{journal}{International Review of Economics \& Finance}\/},  {\it \bibinfo{volume}{84}\/}, \bibinfo{pages}{813--831}.
%Type = Article
\bibitem[{Stevenson et~al.(2021)Stevenson, Mues \& Bravo}]{stevenson2021value}
\bibinfo{author}{Stevenson, M.}, \bibinfo{author}{Mues, C.}, \& \bibinfo{author}{Bravo, C.} (\bibinfo{year}{2021}).
\newblock \bibinfo{title}{The value of text for small business default prediction: A deep learning approach}.
\newblock {\it \bibinfo{journal}{European Journal of Operational Research}\/},  {\it \bibinfo{volume}{295}\/}, \bibinfo{pages}{758--771}.
%Type = Article
\bibitem[{Tabachov{\'a} et~al.(2024)Tabachov{\'a}, Diem, Borsos, Burger \& Thurner}]{tabachova2024estimating}
\bibinfo{author}{Tabachov{\'a}, Z.}, \bibinfo{author}{Diem, C.}, \bibinfo{author}{Borsos, A.}, \bibinfo{author}{Burger, C.}, \& \bibinfo{author}{Thurner, S.} (\bibinfo{year}{2024}).
\newblock \bibinfo{title}{Estimating the impact of supply chain network contagion on financial stability}.
\newblock {\it \bibinfo{journal}{Journal of Financial Stability}\/},  {\it \bibinfo{volume}{75}\/}, \bibinfo{pages}{101336}.
%Type = Article
\bibitem[{Tavakoli et~al.(2025)Tavakoli, Chandra, Tian \& Bravo}]{tavakoli2023multi}
\bibinfo{author}{Tavakoli, M.}, \bibinfo{author}{Chandra, R.}, \bibinfo{author}{Tian, F.}, \& \bibinfo{author}{Bravo, C.} (\bibinfo{year}{2025}).
\newblock \bibinfo{title}{Multi-modal deep learning for credit rating prediction using text and numerical data streams}.
\newblock {\it \bibinfo{journal}{Applied Soft Computing}\/},  {\it \bibinfo{volume}{171}\/}, \bibinfo{pages}{112771}.
%Type = Book
\bibitem[{Thomas et~al.(2017)Thomas, Crook \& Edelman}]{thomas2017credit}
\bibinfo{author}{Thomas, L.}, \bibinfo{author}{Crook, J.}, \& \bibinfo{author}{Edelman, D.} (\bibinfo{year}{2017}).
\newblock {\it \bibinfo{title}{Credit scoring and its applications}\/}.
\newblock \bibinfo{publisher}{SIAM-Society for Industrial and Applied Mathematics}.
%Type = Inproceedings
\bibitem[{Veli{\v{c}}kovi{\'c} et~al.(2018)Veli{\v{c}}kovi{\'c}, Cucurull, Casanova, Romero, Li{\`o} \& Bengio}]{velivckovic2018graph}
\bibinfo{author}{Veli{\v{c}}kovi{\'c}, P.}, \bibinfo{author}{Cucurull, G.}, \bibinfo{author}{Casanova, A.}, \bibinfo{author}{Romero, A.}, \bibinfo{author}{Li{\`o}, P.}, \& \bibinfo{author}{Bengio, Y.} (\bibinfo{year}{2018}).
\newblock \bibinfo{title}{Graph attention networks}.
\newblock In {\it \bibinfo{booktitle}{6th International Conference on Learning Representations (ICLR)}\/}.
%Type = Article
\bibitem[{Vinciotti et~al.(2019)Vinciotti, Tosetti, Moscone \& Lycett}]{vinciotti2019effect}
\bibinfo{author}{Vinciotti, V.}, \bibinfo{author}{Tosetti, E.}, \bibinfo{author}{Moscone, F.}, \& \bibinfo{author}{Lycett, M.} (\bibinfo{year}{2019}).
\newblock \bibinfo{title}{The effect of interfirm financial transactions on the credit risk of small and medium-sized enterprises}.
\newblock {\it \bibinfo{journal}{Journal of the Royal Statistical Society Series A: Statistics in Society}\/},  {\it \bibinfo{volume}{182}\/}, \bibinfo{pages}{1205--1226}.
%Type = Article
\bibitem[{Wang(2016)}]{wang2016biggest}
\bibinfo{author}{Wang, Y.} (\bibinfo{year}{2016}).
\newblock \bibinfo{title}{What are the biggest obstacles to growth of {SME}s in developing countries? -- {A}n empirical evidence from an enterprise survey}.
\newblock {\it \bibinfo{journal}{Borsa Istanbul Review}\/},  {\it \bibinfo{volume}{16}\/}, \bibinfo{pages}{167--176}.
%Type = Article
\bibitem[{Weisfeiler \& Lehman(1968)}]{weisfeiler1968reduction}
\bibinfo{author}{Weisfeiler, B.}, \& \bibinfo{author}{Lehman, A.~A.} (\bibinfo{year}{1968}).
\newblock \bibinfo{title}{A reduction of a graph to a canonical form and an algebra arising during this reduction}.
\newblock {\it \bibinfo{journal}{Nauchno-Technicheskaya Informatsia}\/},  {\it \bibinfo{volume}{2}\/}, \bibinfo{pages}{12--16}.
\newblock \bibinfo{note}{English translation by G.~Ryabov available at \url{https://www.iti.zcu.cz/wl2018/pdf/wl_paper_translation.pdf}}.
%Type = Inproceedings
\bibitem[{Xu et~al.(2019)Xu, Hu, Leskovec \& Jegelka}]{xu2018powerful}
\bibinfo{author}{Xu, K.}, \bibinfo{author}{Hu, W.}, \bibinfo{author}{Leskovec, J.}, \& \bibinfo{author}{Jegelka, S.} (\bibinfo{year}{2019}).
\newblock \bibinfo{title}{How powerful are graph neural networks?}
\newblock In {\it \bibinfo{booktitle}{7th International Conference on Learning Representations (ICLR)}\/}.
%Type = Article
\bibitem[{Yin et~al.(2020)Yin, Jiang, Jain \& Wang}]{yin2020evaluating}
\bibinfo{author}{Yin, C.}, \bibinfo{author}{Jiang, C.}, \bibinfo{author}{Jain, H.~K.}, \& \bibinfo{author}{Wang, Z.} (\bibinfo{year}{2020}).
\newblock \bibinfo{title}{Evaluating the credit risk of {SME}s using legal judgments}.
\newblock {\it \bibinfo{journal}{Decision Support Systems}\/},  {\it \bibinfo{volume}{136}\/}, \bibinfo{pages}{113364}.
%Type = Article
\bibitem[{Zandi et~al.(2025)Zandi, Korangi, {\'O}skarsd{\'o}ttir, Mues \& Bravo}]{zandi2024attention}
\bibinfo{author}{Zandi, S.}, \bibinfo{author}{Korangi, K.}, \bibinfo{author}{{\'O}skarsd{\'o}ttir, M.}, \bibinfo{author}{Mues, C.}, \& \bibinfo{author}{Bravo, C.} (\bibinfo{year}{2025}).
\newblock \bibinfo{title}{Attention-based dynamic multilayer graph neural networks for loan default prediction}.
\newblock {\it \bibinfo{journal}{European Journal of Operational Research}\/},  {\it \bibinfo{volume}{321}\/}, \bibinfo{pages}{586--599}.
%Type = Article
\bibitem[{Zhang et~al.(2020{\natexlab{a}})Zhang, Yang, He \& Deng}]{zhang2020multimodal}
\bibinfo{author}{Zhang, C.}, \bibinfo{author}{Yang, Z.}, \bibinfo{author}{He, X.}, \& \bibinfo{author}{Deng, L.} (\bibinfo{year}{2020}{\natexlab{a}}).
\newblock \bibinfo{title}{Multimodal intelligence: Representation learning, information fusion, and applications}.
\newblock {\it \bibinfo{journal}{IEEE Journal of Selected Topics in Signal Processing}\/},  {\it \bibinfo{volume}{14}\/}, \bibinfo{pages}{478--493}.
%Type = Article
\bibitem[{Zhang et~al.(2020{\natexlab{b}})Zhang, Yin, Zeng, Yuan \& Zhang}]{zhang2020combining}
\bibinfo{author}{Zhang, D.}, \bibinfo{author}{Yin, C.}, \bibinfo{author}{Zeng, J.}, \bibinfo{author}{Yuan, X.}, \& \bibinfo{author}{Zhang, P.} (\bibinfo{year}{2020}{\natexlab{b}}).
\newblock \bibinfo{title}{Combining structured and unstructured data for predictive models: A deep learning approach}.
\newblock {\it \bibinfo{journal}{BMC Medical Informatics and Decision Making}\/},  {\it \bibinfo{volume}{20}\/}, \bibinfo{pages}{1--11}.
%Type = Inproceedings
\bibitem[{Zhang et~al.(2018)Zhang, Xie, Xidao \& Zhang}]{zhang2018multi}
\bibinfo{author}{Zhang, L.}, \bibinfo{author}{Xie, Y.}, \bibinfo{author}{Xidao, L.}, \& \bibinfo{author}{Zhang, X.} (\bibinfo{year}{2018}).
\newblock \bibinfo{title}{Multi-source heterogeneous data fusion}.
\newblock In {\it \bibinfo{booktitle}{2018 International Conference on Artificial Intelligence and Big Data (ICAIBD)}\/} (pp. \bibinfo{pages}{47--51}).
%Type = Article
\bibitem[{Zhang et~al.(2022)Zhang, Yan, Li, Tian \& Yoshida}]{zhang2022credit}
\bibinfo{author}{Zhang, W.}, \bibinfo{author}{Yan, S.}, \bibinfo{author}{Li, J.}, \bibinfo{author}{Tian, X.}, \& \bibinfo{author}{Yoshida, T.} (\bibinfo{year}{2022}).
\newblock \bibinfo{title}{Credit risk prediction of {SME}s in supply chain finance by fusing demographic and behavioral data}.
\newblock {\it \bibinfo{journal}{Transportation Research Part E: Logistics and Transportation Review}\/},  {\it \bibinfo{volume}{158}\/}, \bibinfo{pages}{102611}.
%Type = Article
\bibitem[{Zhao et~al.(2024)Zhao, Zhang \& Geng}]{zhao2024deep}
\bibinfo{author}{Zhao, F.}, \bibinfo{author}{Zhang, C.}, \& \bibinfo{author}{Geng, B.} (\bibinfo{year}{2024}).
\newblock \bibinfo{title}{Deep multimodal data fusion}.
\newblock {\it \bibinfo{journal}{ACM Computing Surveys}\/},  {\it \bibinfo{volume}{56}\/}.
%Type = Article
\bibitem[{Zhou et~al.(2022)Zhou, Lee, Li, Chen, Xie, Wu \& Zeng}]{zhou2022identifying}
\bibinfo{author}{Zhou, T.}, \bibinfo{author}{Lee, Y.-L.}, \bibinfo{author}{Li, Q.}, \bibinfo{author}{Chen, D.}, \bibinfo{author}{Xie, W.}, \bibinfo{author}{Wu, T.}, \& \bibinfo{author}{Zeng, T.} (\bibinfo{year}{2022}).
\newblock \bibinfo{title}{Identifying discreditable firms in a large-scale ownership network}.
\newblock {\it \bibinfo{journal}{arXiv preprint arXiv:2211.14316}\/}, .

\end{thebibliography}

\begin{appendices}
\counterwithin{table}{section}
\counterwithin{figure}{section}
\renewcommand\thetable{\Alph{section}.\arabic{table}}
\renewcommand\thefigure{\Alph{section}.\arabic{figure}}

\newpage

\section{Strategy for handling null values}
\label{Strategy for handling null values}

\begin{table}[hbt!]
\footnotesize
\centering
\caption{Strategy for handling null values.}
\begin{tabular}{cc|l}
\toprule
Null value percentage & Feature type & Action\\
\midrule
\multirow{2}{*}{0\% to $<5\%$} & Categorical & Replace with mode\\
& Numerical & Replace with median\\
\midrule
\multirow{2}{*}{5\% to $<40\%$} & Categorical & Replace with `N/A'\\
& Numerical & Replace with median and add dummy variable indicating presence or absence\\
\midrule
\multirow{2}{*}{40\% to $<95\%$} & Categorical & Drop feature and add dummy variable indicating presence or absence\\
& Numerical & Drop feature and add dummy variable indicating presence or absence\\
\midrule
\multirow{2}{*}{95\% to 100\%} & Categorical & Drop feature\\
& Numerical & Drop feature\\
\bottomrule
\end{tabular}
\end{table}

\section{Hyperparameter tuning for baseline models}
\label{Hyperparameter tuning for baseline models}

\begin{table}[hbt!]
\footnotesize
\caption{Hyperparameter tuning for baseline models.}
\begin{center}
\begin{tabular}{c|l|c}
\toprule
Model & Hyper-parameters & Grid search values\\
\midrule
\multirow{1}{*}{LR} & Penalty & \{L1, L2\}\\
\midrule
\multirow{3}{*}{RF} & Number of trees & \{100, 200, 500\}\\
& Minimum samples leaf & \{2, 5, 10\}\\
& Minimum samples split & \{2, 5, 10\}\\
\midrule
\multirow{4}{*}{XGB} & Learning rate & \{0.001, 0.01, 0.1\}\\
& Maximum depth & \{2, 3, 4\}\\
& Number of estimators & \{50, 100, 250, 500\}\\
& Alpha & \{0.1, 0.2, 0.3, 0.4, 0.5, 0.6, 0.7, 0.8, 0.9\}\\
\bottomrule
\end{tabular}
\end{center}
\end{table}

\section{Architecture of the DNN baseline model}
\label{Architecture of the DNN baseline model}

\begin{figure}[hbt!]
\includegraphics[scale=0.58]{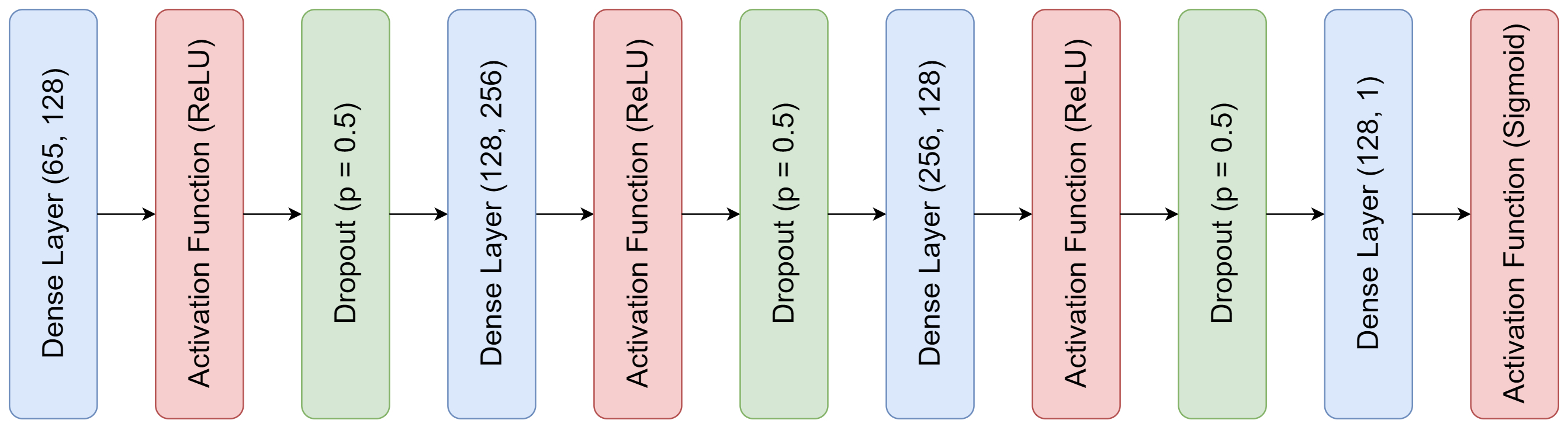}
\centering
\caption{Architecture of the DNN baseline model.}
\label{fig11}
\end{figure}

\section{Hyperparameter tuning for GAT and GIN models}
\label{Hyperparameter tuning for GAT and GIN models}

\begin{table}[hbt!]
\footnotesize
\caption{Hyperparameter tuning for GAT and GIN models.}
\begin{center}
\begin{tabular}{c|l|c}
\toprule
Model & Hyper-parameters & Grid search values\\
\midrule
\multirow{5}{*}{GAT} & Number of layers & \{1, 2\}\\
& Number of attention heads & \{1, 2, 4\}\\
& Hidden units per head & \{16, 32, 64\}\\
& Learning rate & \{0.001, 0.005, 0.01\}\\
& Dropout rate & \{0.0, 0.25, 0.5\}\\
\midrule
\multirow{5}{*}{GIN} & Number of layers & \{1, 2\}\\
& Hidden units & \{16, 32, 64\}\\
& Epsilon (\(\varepsilon\)) & \{0, 0.2, 0.4\}\\
& Learning rate & \{0.001, 0.005, 0.01\}\\
& Dropout rate & \{0.0, 0.25, 0.5\}\\
\bottomrule
\end{tabular}
\end{center}
\end{table}
\end{appendices}

\end{document}